\documentclass[conference]{IEEEtran}
\usepackage{cite}
\usepackage{amsmath,amssymb,amsfonts}
\usepackage{graphicx}
\usepackage{textcomp}
\usepackage{xspace}
\usepackage{color}
\usepackage{subcaption}
\usepackage{hyperref}
\usepackage{booktabs}
\usepackage{mdframed}
\usepackage{algorithm}
\usepackage{algorithmicx}
\usepackage{algpseudocode}

\newcommand{\codename}{\textsf{VIF}\xspace}
\newcommand{\name}{\codename}

\newcommand{\updated}[1]{{#1}}
\newcommand{\deleted}[1]{}

\newenvironment{packeditemize}{
\begin{list}{$\bullet$}{
\setlength{\itemsep}{1.5pt}
\setlength{\labelwidth}{8pt}
\setlength{\leftmargin}{10pt}
\setlength{\labelsep}{3pt}
\setlength{\listparindent}{\parindent}
\setlength{\parsep}{1.5pt}
\setlength{\parskip}{1.5pt}
\setlength{\topsep}{1.5pt}}}{\end{list}}

\renewcommand{\paragraph}[1]{\par\noindent\textbf{#1}}

\newcounter{packednmbr}

\begin{document}

\title{
{\huge Practical Verifiable In-network Filtering for DDoS defense}
}

\author{
\IEEEauthorblockN{Deli Gong\IEEEauthorrefmark{1},
Muoi Tran\IEEEauthorrefmark{1}, 
Shweta Shinde\IEEEauthorrefmark{2},
Hao Jin\IEEEauthorrefmark{3},
Vyas Sekar\IEEEauthorrefmark{4}, 
Prateek Saxena\IEEEauthorrefmark{1}, 
Min Suk Kang\IEEEauthorrefmark{1}}

\IEEEauthorblockA{\IEEEauthorrefmark{1}National University of Singapore, \emph{\{gongdeli, muoitran, prateeks, kangms\}@comp.nus.edu.sg}}
\IEEEauthorblockA{\IEEEauthorrefmark{2}University of California, Berkeley, \emph{shwetasshinde24@gmail.com}}
\IEEEauthorblockA{\IEEEauthorrefmark{3}Texas A\&M University, \emph{haojin@tamu.edu}}
\IEEEauthorblockA{\IEEEauthorrefmark{4}Carnegie Mellon University, \emph{vsekar@andrew.cmu.edu}}
}

\maketitle

\begin{abstract}
In light of ever-increasing scale and sophistication of modern DDoS attacks, it is time to revisit {\em in-network filtering} or the idea of empowering DDoS victims to install in-network traffic filters in the upstream transit networks. 
Recent proposals show that filtering DDoS traffic at a handful of large transit networks can handle volumetric DDoS attacks effectively. 
However, the in-network filtering primitive can also be misused.
Transit networks can use the in-network filtering service as an excuse for any arbitrary packet drops made for their own benefit.
For example, transit networks may intentionally execute filtering services poorly or unfairly to discriminate their competing neighbor ASes while claiming that they drop packets for the sake of DDoS defense.
We argue that it is due to the {\em lack of verifiable filtering}\,---\,i.e., no one can check if a transit network executes the filter rules correctly as requested by the DDoS victims. 
To make in-network filtering a more robust defense primitive, in this paper, we propose a verifiable in-network filtering, called \name, that exploits emerging hardware-based trusted execution environments (TEEs) and offers filtering verifiability to DDoS victims and neighbor ASes.
Our proof of concept demonstrates that a \name filter implementation on commodity servers with TEE support can handle traffic at line rate (e.g., 10 Gb/s) and execute up to 3,000 filter rules.  
We show that \name can easily scale to handle larger traffic volume (e.g., 500 Gb/s) and more complex filtering operations (e.g., 150,000 filter rules) by parallelizing the TEE-based filters.
As a practical deployment model, we suggest that Internet exchange points (IXPs) are the ideal candidates for the early adopters of our verifiable filters due to their central locations and flexible software-defined architecture.  
Our large-scale simulations of two realistic attacks (i.e., DNS amplification, Mirai-based flooding) show that only a small number (e.g., 5--25) of large IXPs are needed to offer the \name filtering service to handle the majority (e.g., up to 80--90\%) of DDoS traffic. 
\end{abstract}

\vspace{-5pt}
\section{Introduction}
\label{sec:intro}
\vspace{-5pt}


Distributed denial-of-service (DDoS) attacks are highly prevalent. 
In the last decade, new attack strategies such as amplification~\cite{rossow2014amplification} and new attack sources such as IoT devices~\cite{guardian2016ddos} have surfaced, which have resulted in attacks of extremely high volume~\cite{zdnet2018ddos}.

A large number of  DDoS defenses have been extensively studied over the past two decades.
Among them, an effective defense against the ever-increasing scale of DDoS attacks is {\em in-network filtering} or empowering DDoS victim networks to install in-network traffic filters in the upstream transit networks. 
This idea was proposed in early efforts (e.g., Pushback~\cite{mahajan2002controlling},
D-WARD~\cite{mirkovic2002attacking}, AITF~\cite{argyraki2005active}) and has repeatedly re-surfaced in standardization committees (e.g., see the recent DDoS Open Threat Signaling~\cite{dots-slides}). 

%


Dropping suspicious packets closer to the attack sources \updated{at the requests of DDoS victims} is desirable because
it (1) reduces wasted traffic on downstream ISPs, thereby reducing overall
network bandwidth and cost of routing malicious traffic; and (2) has the potential to
handle ever-increasing attack volume (e.g., several Tb/s) since attack volume
at each distributed filtering point can be much lower than the aggregated volume at the victim~\cite{argyraki2005active}. 

Unsurprisingly, there is renewed interest in the community on revisiting in-network filtering solutions. Indeed, a recent DDoS defense architecture, called SENSS~\cite{ramanathan2018senss}, suggests that the traffic filters installed at a few large transit ISPs directly by the remote DDoS victims can prevent most of the volumetric attack traffic from flooding the victim networks.

However, the in-network filtering primitive can be unfortunately misused by malicious transit networks.
Transit networks may claim to provide in-network filtering service on behalf of a remote DDoS victim only to justify any arbitrary packet drops executed for their own benefit.
For example, a transit network that offers in-network filtering service (which we call a {\em filtering network}) can easily compromise a filter rule {\tt [Drop 50\% of HTTP flows destined to the victim network]} requested by a DDoS victim to harm its neighbor networks, as illustrated in Figure~\ref{fig:venef}.
First, the malicious filtering network can silently discriminate its neighbor upstream ASes by applying arbitrarily modified filter rules (e.g., dropping 80\% of HTTP traffic from AS $A$ but only 20\% from AS $B$) based on its business preference (say, AS $A$ is AT\&T and $B$ is Comcast).
Discrimination and dispute among transit ISPs are not a new problem (e.g., a dispute between Level3 and Comcast~\cite{dispute_level3_comcast}); however, it can be easily exacerbated if transit networks start dropping packets for the sake of DDoS defense. 
Worse yet, the filtering network can also perform the requested filter rules inaccurately (e.g., drop only 20\% HTTP traffic on average) to save its filtering resources while claiming the faithful execution of the requested rules.

We argue that the {\em verifiability} of in-network filtering mitigates such misbehaviors 
by malicious filtering service providers. 
With filtering verifiability, when a filtering network executes a manipulated filter rule, the DDoS victim network who requested the filter rule or the direct neighbor autonomous system (ASes) of the filtering network can detect the misbehavior immediately.

In this paper, we offer the first technical means of the strong verifiability of filtering operations and develop a practical and scalable in-network filtering system, called \name.\footnote{\name stands for `Verifiable In-network Filtering'.}
\name has a generic architecture designed for any transit networks (e.g., Tier-1, large Tier-2 ISPs, IXPs).
We exploit software networking functions running on commodity hardware with trusted execution environments (TEEs), such as Intel SGX~\cite{costan2016intel} and Arm TrustZone~\cite{arm2009security}, which are widely available in commercial off-the-shelf (COTS) server platforms; see Microsoft and Google's SGX-based cloud platforms~\cite{russinovich2017azure-sgx,porter2018asylo-sgx}. 

We identify two key technical challenges to realizing the \name vision in practice: 

\begin{figure}[t!]
	\centering
	\includegraphics[width=0.4\textwidth]{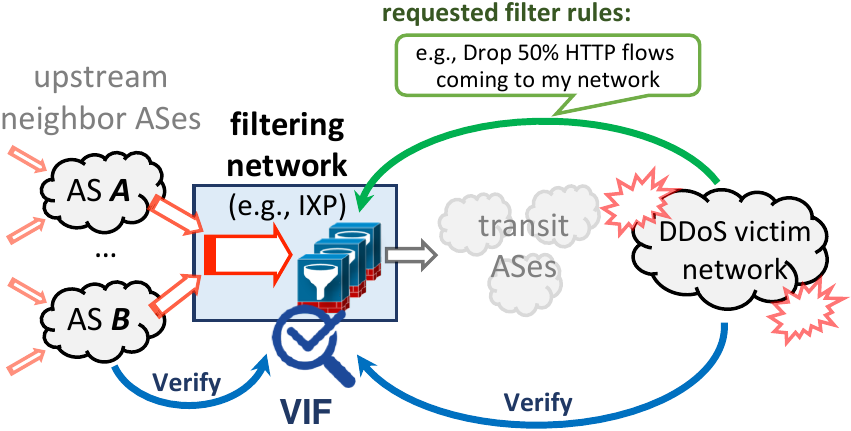}
	\caption{Example of in-network filtering in a transit network. \name enables the direct upstream neighbors/the DDoS victim to {\em verify} if the filtering network executes the requested filter rules correctly.}
	\label{fig:venef}
\end{figure}

\begin{packeditemize}
	\item {\bf Auditability.} To make the in-network filtering verifiable, we exploit the trusted execution environment with hardware-based root of trust (e.g., Intel SGX) and provide integrity guarantee for filter logic. 
	Yet, achieving auditable filtering operations is not trivial.
	One challenge is that the correct filtering operations are also affected by {\em external} inputs to the filter (e.g., incoming packet order, time clock feeds), which can be controlled by a malicious filtering network.
	Another challenge is that a malicious filtering network can still {\em bypass} the auditable filters by reconfiguring its network and avoid using the auditable filters for adversary-selected packets. 
	\item {\bf Scalability.} To scale out the filter capacity for large volumetric attacks, parallelization with multiple TEE-based auditable filters is required; however, some necessary network components for parallelization (e.g., traffic load balancers) are not directly auditable.
	Moreover, distributing filter rules across multiple auditable filters creates an optimization challenge, which involves two dimensions of resource constraints. 
\end{packeditemize}

\paragraph{Approach and Contributions.} 
\name's design makes the following key contributions to address these aforementioned challenges:

\begin{packeditemize}
\item  We analyze the requirements for auditable filters, particularly, their reliance on the external inputs to the filters, which can be controlled by malicious filtering networks (\S\ref{sec:stateless-filter-design}).
Our key insight is that it is sufficient that traffic filters are {\em stateless} for auditable filter operations.
We also implement an effective {\em bypass detection} that relies on the accountable packet logs (with an efficient sketch implementation) measured inside the TEE (\S\ref{sec:filter-bypass-detection}). 
The packet logs can be used to identify packet drops/injections made outside of the auditable filters. 
We demonstrate that an efficient {\em line-rate} implementation (nearly 10 Gb/s throughput performance) of the auditable traffic filters with the TEE support is possible with several system optimizations (\S\ref{sec:eval}); and

\item For highly scalable filtering architecture, we implement a dynamic filter rule distribution algorithm across multiple auditable filters and untrusted network components~\cite{patel2013ananta,gandhi2015duet} (\S\ref{sec:scalable-filter-design}). 
We implement a heuristic that can quickly reconfigure a large number of filter rules (e.g., 150,000 filter rules) and a large volume of incoming traffic (e.g., the total volume of 500 Gb/s) with auditable filter instances (e.g., 50 filters). 

\end{packeditemize}

As an early deployment model, we suggest to deploy \name in major Internet exchange points (IXPs) (\S\ref{sec:deployment-proposals}).
In the last decade, IXPs have become the central infrastructure of the global Internet connectivity~\cite{ager2012anatomy,chatzis2013there}, with large IXPs handling daily traffic volumes comparable to those carried by the largest Tier-1 ISPs, which makes the IXPs the perfect candidates for our verifiable filtering service~\cite{dietzel2016blackholing}.
We perform large-scale inter-domain simulations with two realistic attacks (i.e., DNS amplification, Mirai-based DDoS attacks) and show that deploying \name in a small number (e.g., 5--25) of large IXPs is enough to handle the majority (e.g., up to 90\%) of DDoS attacks (\S\ref{subsec:against-ddos-exp}). 

\section{Problem Definition}
\label{sec:problem-definition}


In this section, we describe the threat model we consider in this paper (\S\ref{subsec:threat-model}), the desired properties of the \name design, the trusted execution environment model (\S\ref{subsec:trusted-execution-environment}) (\S\ref{sec:desired-filter-verifiability}) and the assumption we make in this work (\S\ref{sec:assumptions}).

\subsection{Threat Model}
\label{subsec:threat-model}

We consider a single malicious transit network that offers in-network filtering services for downstream networks.
In this threat model, the adversary has full control of the control plane and data plane within its network.

Our threat model focuses on the problem of the malicious filtering network manipulating the DDoS-victim-submitted filter rules, which we call as {\em filter rule violation attacks}.
In general, let us consider a filter rule $R$ requested by a victim network.
The malicious filtering network may change $R$ arbitrarily into a different filter rule $R'$ and apply it to all traffic destined to the victim network.
Also, the malicious filtering network may apply different modified rules $R'_1, R'_2, \cdots$ for different traffic flows (e.g., packets delivered via different neighboring ASes).
In the followings, we present two example attack goals, where the manipulation of filter rules at a filtering network can seriously disrupt the packet forwarding services for the neighboring ASes and the remote DDoS victim networks.
In both attacks, we use the same example as in Section~\ref{sec:intro}, i.e., the filtering network manipulates the DDoS-victim-submitted filter rule $R=${\tt [Drop 50\% of HTTP flows destined to victim network]}.

\paragraph{[Goal 1] Discriminating neighboring ASes.}
ASes expect their traffic to be reliably forwarded by the transit networks.
Yet, when a transit network offers the filtering services for the sake of DDoS defense, the packet forwarding service can be silently degraded differently for different neighboring upstream ASes. 
%
In particular, the filtering network can apply the {modified} filter rules for each neighboring AS based on its own business preference.
Instead of applying the same original rule $R$, the filtering network may apply $R'_A=${\tt [Drop 20\% of HTTP flows destined to victim network]} for traffic flows delivered by AS $A$ and $R'_B=${\tt [Drop 80\% of HTTP flows destined to victim network]} for traffic flows delivered by AS $B$. 
Such discriminatory filtering is hard to detect because individual neighbor AS $A$ and $B$ do not know the unmodified rule $R$ and, even if they know $R$, they cannot determine if the traffic filter applied to their packets is $R$. 
Each neighbor AS may try to infer the packet drops of the end-to-end path indirectly (e.g., via monitoring TCP sessions); yet, it is {\em insufficient} because pinpointing exact location of packet losses (a.k.a. fault localization) is known to be hard without large-scale network collaboration~\cite{argyraki2010verifiable, basescu2016high}.



\paragraph{[Goal 2] Reducing operational cost with inaccurate filtering.}
%
To reduce the operational cost, the malicious filtering network may violate the filter rules submitted by the DDoS victims. 
%
For instance, consider that the malicious filtering network wants to use {\em only} 10 Gb/s of its filtering capacity for the rule $R$ while total incoming traffic of 50 Gb/s should be sent to the filters.
To achieve this, the filtering network can send only 10 Gb/s traffic to its filters and execute the unmodified rule $R$.
For the rest of the 40 Gb/s traffic, however, the filtering network can simply allow or drop all
{\em without} using the filtering capacity.
Since the victim network has no information about the incoming traffic arrived at the filtering network, it cannot directly detect this attack. 




\subsection{Desired Properties: Filtering Verifiability and Scalability}
\label{sec:desired-filter-verifiability}

As we introduced in the Introduction, we have the two desired properties of the \codename design. 
First, to remove the attack capabilities required for the filter-rule violation attacks, \codename desires to have the {\em filtering verifiability}.
A filter rule $R$ is said to be verifiable if any modification of the rule $R$ and its execution by the filtering network can be detected by the victim networks or the direct neighboring ASes.
%
%
Second, our \name system must be easily {\em scaled up} because DDoS attacks are increasingly scalable in both volume and complexity. 
We have observed an escalation in the volume of DDoS attack
traffic~\cite{computerworld2013biggest,guardian2016ddos} and attacks are getting more sophisticated; e.g., multi-million bots are becoming more common~\cite{mirai}. 

\subsection{Trusted Execution Environment with Intel SGX}
\label{subsec:trusted-execution-environment}
\codename uses TEEs, particularly Intel SGX in this work, as the feasible hardware-based root of trust.
Intel Software Guard Extensions (SGX) is a recent architectural feature that allows secure execution
of a program on a computing infrastructure in control of
an adversarial operator~\cite{costan2016intel,mckeen2016intel}.
SGX also supports secure execution of a user-level program with no modification of underlying commodity software stacks~\cite{shinde2017panoply,baumann2014shielding,arnautov2016scone}. 
In particular, it offers the isolated execution of the application logic in a protected memory region, called an {\em enclave}, which prevents the operator from tampering it.
Moreover, it supports remote attestation that allows a third party to audit that the correct application and data has been loaded in an enclave.
%
The attestation process starts when a verifier issues an attestation challenge to the enclaved application. 
The enclave then provides a report, which is cryptographically signed with the attestation key of the SGX hardware. 
Next, the attestation report is verified by the Intel Attestation Service (IAS), which is distributed globally~\cite{johnson2016intel}.
Recently, Intel also allows {\em anyone}, who gets a certificate from Intel, to run their own remote attestation services and verify the attestation report~\cite{vinnie2018intel}.

\subsection{Assumptions} 
\label{sec:assumptions}
We assume that ISPs (e.g., victim networks) trust the remote attestation process for the integrity guarantees of the \name enclave. 
We also assume an idealized implementation of \codename that has no backdoor.
The full implementation of \codename is open-sourced for public scrutiny\footnote{\name is open sourced at \url{https://github.com/InNetworkFiltering/SGX-DPDK}}.
We leave a formal verification of \codename implementation as future work.
We consider all hardware and side-channel attacks (e.g.,\cite{gotzfried2017cache, last-level-cache, hahnel2017high}) as out of scope of this paper since countermeasures to these (e.g.,~\cite{chen2017detecting,gruss2017strong,shih2017t,pigeonhole,costan2016sanctum,song2018towards}) are orthogonal to the design of \codename. 

\section{Auditable Filter Design}
\label{sec:filter-design}

The \codename filtering operation is enclosed by an SGX enclave where the integrity of its execution is guaranteed, i.e., a malicious filtering network cannot tamper it.
Furthermore, the filtering internal logic and states are also securely verified via the remote attestation process~\cite{johnson2016intel}.
The isolated execution and remote attestation are useful in realizing the auditable filter; yet, they are insufficient because (1) the filtering decisions can be influenced by the external inputs to the filter such as packet order and time clock feeds, which are controlled by the filtering network; and (2) the malicious filtering network may redirect the traffic within its network to bypass the filtering operations.

To address these two challenges, we propose the filtering operations to be {\em stateless} and hence be independent from the external inputs (\S\ref{sec:stateless-filter-design}) and implement the enclaved packet logs to detect bypassing attempts (\S\ref{sec:filter-bypass-detection}).




\subsection{Stateless Filter Design}
\label{sec:stateless-filter-design}

To analyze the dependencies of the enclaved filters, let us first describe an abstract model for our enclaved filter $f$: 
\begin{align}
\{\texttt{ALLOW},\texttt{DROP}\}\leftarrow f\big(\langle{p,a}\rangle, (\langle{p_1,a_1}\rangle, \langle{p_2,a_2}\rangle, \cdots)\big),
\end{align}
where $\langle{p_i,a_i}\rangle$ denotes that packet $p_i$ arrives at the enclaved filter at time $a_i$ (measured by the enclave's internal clock), $\langle{p,a}\rangle$ represents the packet $p$ that is being evaluated and its arrival time $a$, and the following time relationship holds $a > a_1 \ge a_2 \ge \cdots$.



Notice that in this abstract model, the filtering operation of a packet $p$ depends on the packet arrival time and the order of the packets, which can be exploited by the filtering network (see next).
Here, we summarize the two properties that are needed to make \codename filter auditable:
\begin{packeditemize}
    \item {\em Arrival-time independence.} The 
    filtering decision should be independent of packet-arrival time because it can be easily manipulated by a malicious filtering network (e.g., delaying individual data packets). 
    Moreover, a malicious filtering network can delay the time query/response messages to/from the trusted clock source for the enclave~\cite{sgx-forum-trusted-time}, slowing down the enclave's internal time clock.
    \item {\em Packet-injection independence.}
    The filtering decision should not depend on the previous packets since a malicious filtering network can also inject any arbitrary packets into the traffic flow and influence the filtering decision.
\end{packeditemize}

Thus, to ensure that the filtering operations are auditable, the filtering function $f$ can be {\em independent} of all the previous packets and their arrival times; that is,
\begin{align}
f\big(\langle{p,a}\rangle, (\langle{p_1,a_1}\rangle, \langle{p_2,a_2}\rangle, \cdots)\big)=f(p),
\end{align}
which simplifies the filter design to $n$-tuple (e.g., {\tt srcIP, dstIP, srcPort, dstPort, protocol}) per-packet filters.
In other words, the filtering decision of packet $p$ solely relies on $p$, such as five-tuple bits.
While we acknowledge the limitations of such a simple $n$-tuple filter design (e.g., incapable of handling complicated application-level DoS attacks), we note that it is sufficient for handling large volumetric attacks that are in imminent need of in-network filtering.


Particularly, for handling volumetric attacks, we allow victim networks to express filter rules for exact-match five-tuple flows (e.g., a specific TCP flow between two hosts) or coarse-grained flow specifications (e.g., HTTP connections from hosts in a /24 prefix).
Details for several practical design points for our enclaved filter are found in Appendix~\ref{sec:filter-design-points}.

\subsection{Filter Bypass Detection}
\label{sec:filter-bypass-detection}

The auditability of the \name filter guarantees that the filter operates correctly for the given packets from a filtering network to a victim network.
However, packets may not be properly filtered when a malicious filtering network configures the traffic to {\em bypass} the \name filter, hence violating the filter rules.
Particularly, the manipulation of the traffic flows happen {\em outside} of the protected enclave and thus cannot be detected by the auditable \name filter itself.
We categorize the filter bypass attacks as follows:
\begin{packeditemize}
	\item {\em Injection after filtering:} The \name filter drops a packet $p$ but the adversary injects a copy of $p$ into the packet stream that is forwarded to the victim;
	\item {\em Drop after filtering:} A packet $p$ is allowed by the filter but the adversary drops $p$ before forwarding it to the victim; and 
	\item {\em Drop before filtering:} The filtering network drops a packet $p$ even before it is processed by the filter.
\end{packeditemize}
%
%
%
%
  
Note that, we do {\em not} consider {\em injection before filtering} operations by a filtering network as an attack because it does not affect the filtering decision due to the {\em packet-injection independence} property of the filters (see Section~\ref{sec:stateless-filter-design}).\footnote{Moreover, the detection of packet injections before the enclave operations is hard without explicit coordination with traffic sources.}

\begin{figure}[t!]
	\centering
	\includegraphics[width=0.45\textwidth]{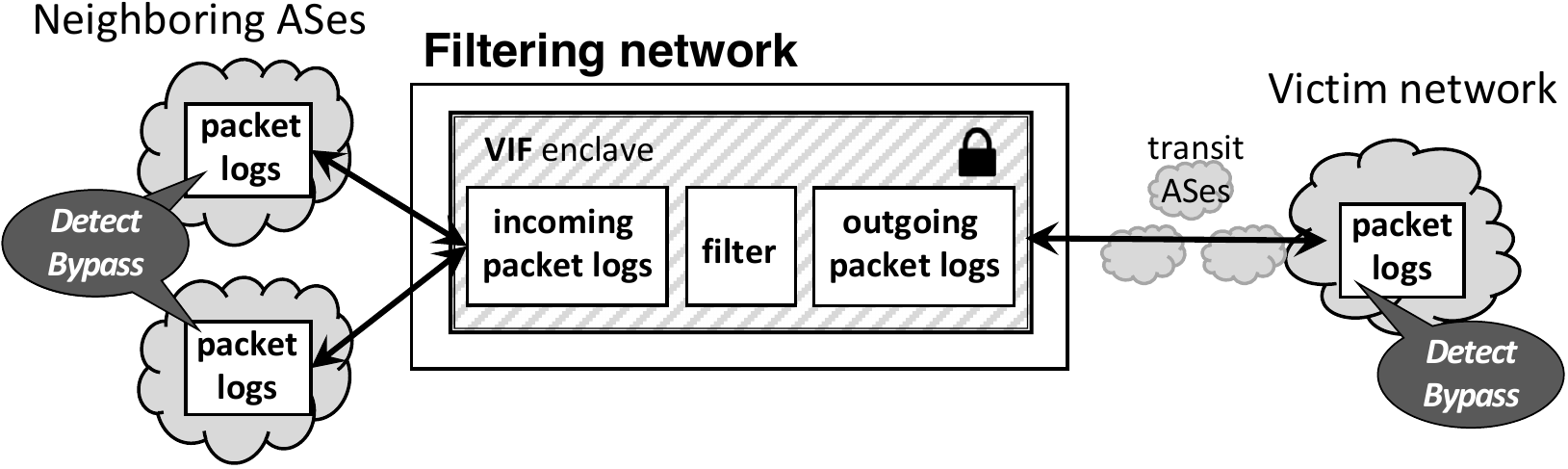}
	\caption{Bypass detection. Efficient sketch data structure is used for packet logs.}
	\label{fig:bypass-detection}
\end{figure}

\paragraph{Bypass Detection.}
%
We allow the victim network and the neighboring ASes of the filtering network to detect such bypass attempts by implementing the {\em accountable packet logs} inside the enclave for incoming and outgoing packet streams, see Figure~\ref{fig:bypass-detection}.
For each packet log, we utilize a sketch, particularly a {\em count-min sketch}, a memory-efficient data structure that stores summaries of streaming data~\cite{cormode2005improved}.
With the sketch-based packet logging, the \name filter keeps only the measurement summary inside an enclave and significantly minimize the memory footprint; e.g., less than 1 MB per each sketch. 
With some additional data-plane optimizations (see Section~\ref{subsec:implementation}), the computational overhead of computing two sketches per packet is negligible (see Section~\ref{sec:data-plane-performance}). 

To detect bypass attempts by the filtering network, the victim network queries the authenticated outgoing packet logs from the \codename enclave and compares it with its own local sketch. 
Any discrepancy between the two sketches implies {\em injection after filter} and/or {\em drop after filter} attacks by the filtering network. 
The computation and bandwidth overhead for the logs queries is negligible; i.e., sketching is highly efficient and requires sending only a few MB of sketch memory via the already established channel with the victim network.~\footnote{The computational overhead of the victim network should also be low since it only requires an efficient sketch on a commodity server without SGX overhead.}
When the victim network detects any bypass attempt, it can decide to abort the ongoing filtering request with the filtering network.
In practice, the \name filtering network should allow a short (e.g., a few minutes) time duration for each filtering round so that victim networks can abort any further request quickly when it detects any bypass attempts. 

Similarly, individual neighbor ASes of the filtering network can detect the {\em drop before filtering} attacks by comparing their own local packet logs with the authenticated incoming packet logs of the \name filter.
%
As a result, the neighboring ASes can choose another downstream network when they obtain the evidence that their downstream network offers filtering services but intentionally drops their packets before they reach the \name filters. 

\paragraph{Handling malicious intermediate ASes.}
The bypass detection mechanism may cause {\em false positives} when some packets are dropped after leaving the \name filter but before reaching the victim network. 
When this happens, the victim network cannot accurately pinpoint where the packet drop has happened. 
The packet could have been dropped by one of the {\em intermediate} transit networks between the \name filtering network and the victim network, or by the \name filtering network itself. 
This, often-called, fault localization is known to be difficult unless all the networks (including the \name filtering network, the victim network, and intermediate ASes) collaborate~\cite{argyraki2010verifiable,basescu2016high}.
However, such large-scale collaboration is unlikely in today's Internet. 

Therefore, instead of locating such packet drops, \name allows the victim network to dynamically {\em test} all the intermediate ASes by rerouting its inbound traffic to avoid each of ASes being tested in a short time using the well-known BGP poisoning-based techniques (e.g., \cite{katz2012lifeguard,smith2018routing}).
Due to the space constraint, we describe the detailed test steps in Appendix~\ref{app:dynamic-test-intermediate-ases}.

\section{Scalable Filter Design}
\label{sec:scalable-filter-design}


In this section, we first analyze the performance (e.g., throughput, network I/O) bottlenecks of a single auditable filter (\S\ref{sec:bottleneck-rule}) and then describe a scalable filtering design with multiple enclaved filters running in parallel and an untrusted load balancer (\S\ref{sec:optimization-multiple-enclaves}).

\subsection{Bottlenecks: Maximum Bandwidth and Number of Rules per SGX Filter}
\label{sec:bottleneck-rule}

\begin{figure}[t]
    \centering
    \begin{subfigure}[t]{0.22\textwidth}
        \centering
        \includegraphics[width=\textwidth]{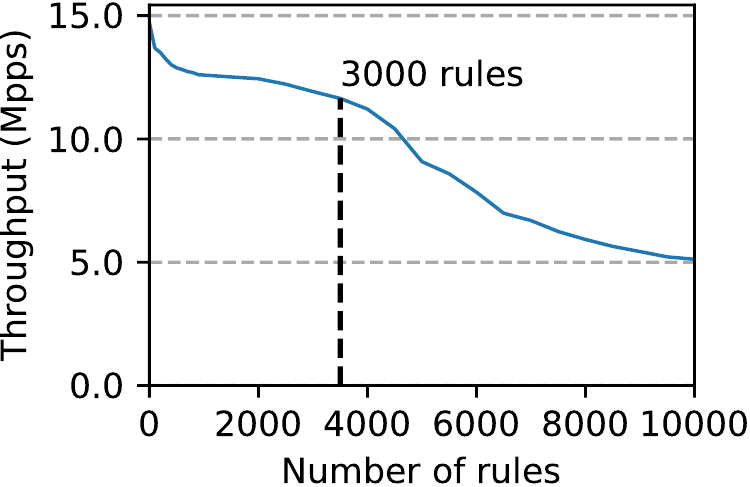}
        \caption{Filter throughput}
        \label{fig:thruput-bottleneck}
    \end{subfigure}
    \begin{subfigure}[t]{0.22\textwidth}
        \centering
        \includegraphics[width=\textwidth]{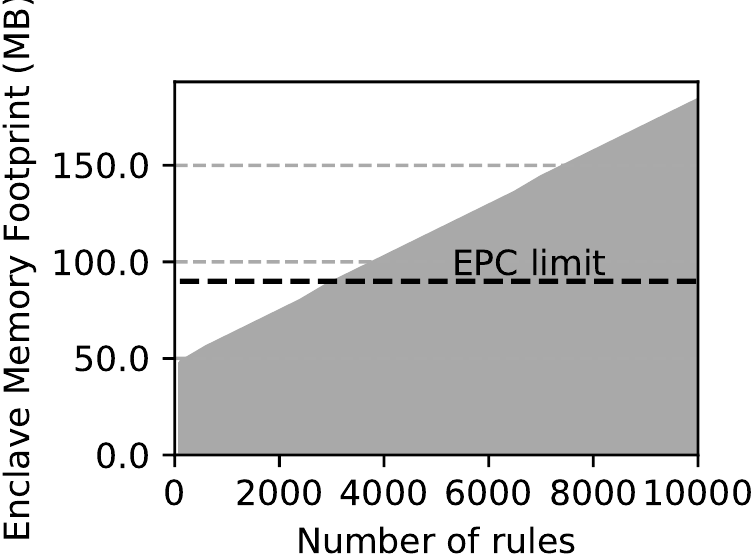}
        \caption{Filter memory footprint}
        \label{fig:mem-bottlenck}
    \end{subfigure}
    \caption{Filter throughput degradation for increasing number of filtering rules}
    \label{fig:bottleneck}
\end{figure}

Recent works such as mbTLS have demonstrated the 10Gb/s performance per enclave can be reached with a four SGX cores machine~\cite{naylor2017and}.
Although the processors with six or more cores available on the market~\footnote{List of SGX-enabled processors is available at: \url{ark.intel.com}.} may support larger bandwidth, we consider the maximum network I/O performance of each SGX enclave is 10 Gb/s in the rest of the paper.

Since the SGX-based filter must match the installed rules with incoming flows to perform filtering, the number of filter rules naturally becomes the bottleneck of the filter's performance.
Indeed, we measure the throughput of traffic processed by a single enclaved filter with different numbers of filter rules and show the results in Figure~\ref{fig:thruput-bottleneck}.
We can see from Figure~\ref{fig:thruput-bottleneck} that when the number of filter rules exceeds approximately 3,000, the \name filter's throughput performance rapidly degrades.

One of the explanation is that, when the number of filter rules increases, the lookup table for the packet processing inside an SGX enclave also grows accordingly. 
Even when we use the state-of-the-art multi-bit tries data structure for looking up the filter rules (see Section~\ref{sec:eval} for details), the memory size of the lookup table still grows linearly with the number of filter rules, as shown in Figure~\ref{fig:mem-bottlenck}.
This result also confirms the Enclave Page Cache (EPC) limit is around 92 MB, as seen in many other works (e.g.,~\cite{kuvaiskii2018snort}).



\subsection{Scalable Filtering with Multiple SGX Filters}
\label{sec:optimization-multiple-enclaves}


\begin{figure}[t!]
	\centering
	\includegraphics[width=0.45\textwidth]{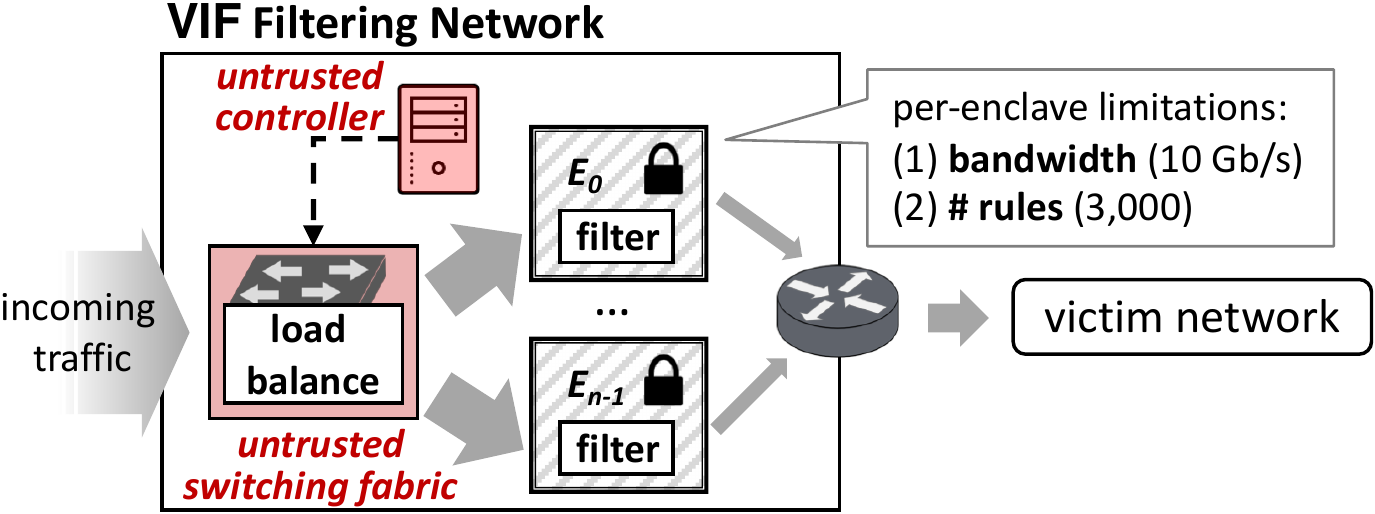}
	\caption{Scalable \name architecture. Multiple \name enclaves are parallelized with an untrusted load balancer.}
	\label{fig:scalability}
\end{figure}

Given the architectural limitation of secure computing resources in currently available SGX architecture, the single-enclave filtering deployment may not be able to deal with the increasing attack volume and complexity.
Hence, we propose a generic \codename architecture that can easily scale up as the number of filters grows, as shown in Figure~\ref{fig:scalability}.
The scalable \codename design includes multiple enclaved filters running in parallel and some {\em untrusted} facilitating components such as the high-bandwidth switching fabric and the controller.

In the followings, we describe how filter rules are distributed and dynamically adjusted among multiple enclaves and explain why a malicious filtering network cannot exploit the untrusted components to manipulate the filtering operations.

\begin{figure}[t!]
	\centering
	\includegraphics[width=0.30\textwidth]{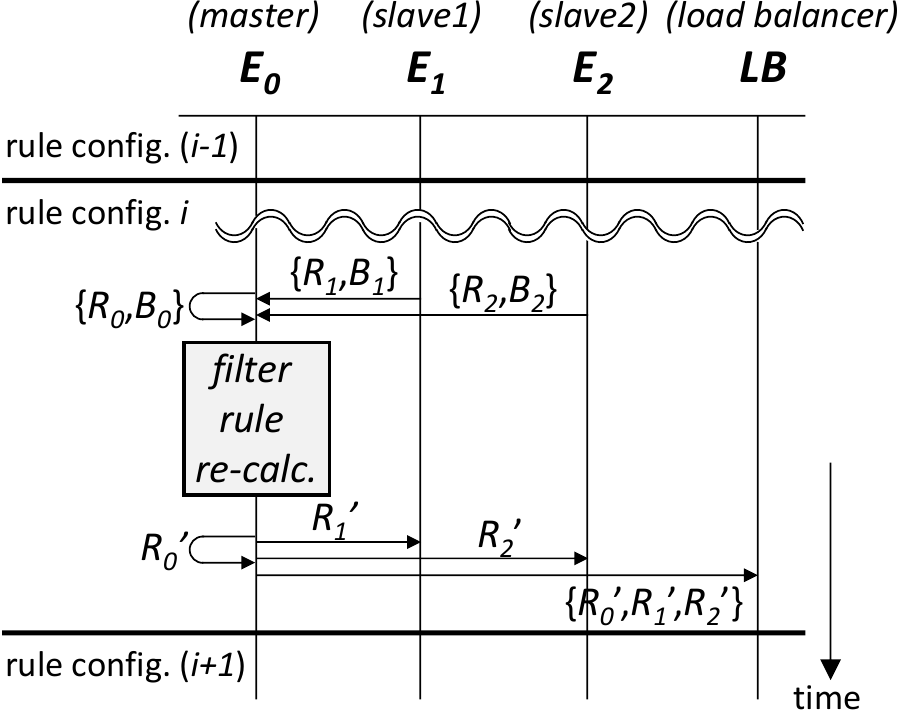}
	\vspace{-5pt}
	\caption{Protocol for filter rule recalculation and redistribution across three enclaved filters: $E_0$, $E_1$, and $E_2$.}
	\vspace{-10pt}
	\label{fig:multi-enclave-protocol}
\end{figure}

\paragraph{Filter rules distribution protocol.}
Since the traffic flows being filtered are frequently changed, the filter rules also need to be updated and redistributed among filters accordingly.
We consider the filter rules distribution happens in {\em rounds}, i.e., the entire filter rule set is given and does not change until the next rule reconfiguration is executed.
In each round, the filter rules are calculated and redistributed via a simple master-slave topology among multiple enclaved filters.
We illustrate the protocol in Figure~\ref{fig:multi-enclave-protocol}, where we have filter $E_0$ as the master node and $E_1$, $E_2$ as the slave nodes.
In particular, when reconfiguration of filter rules is desired (e.g., traffic volume or the number of filter rules handled by a certain filter exceeds a threshold), any enclaved filter may initiate a rule redistribution round and become the master node.
Then, all the slave nodes upload their filter rule sets ($R_i$ for $E_i$) and the array of the average received flow rates of each rule set $R_i$ ($B_i$ for $E_i$) to the master node.
The master node calculates re-configured filter rules, which then be redistributed to all the slave nodes and the load balancer. 
If the calculation requires the changes to the number of enclaves, necessary additional steps (e.g., creating and attesting more enclaved filters) may be required before the rule redistribution.

\paragraph{Filter rules calculation optimization problem.} 
In each filter rules redistribution, the master node has to allocate the bandwidth and rules to all enclaved filters.
We consider the calculation of the optimal rule sets for each filter enclave as solving a mixed integer linear programming (ILP) optimization problem. 
We assume $k$ filter rules as $r_i$ ($1 \leq i \leq k$) and the corresponding incoming bandwidth as $b_i$ ($1 \leq i \leq k$).\footnote{We denote $b_i$ as the incoming bandwidth measured for a filter rule for easier understanding. 
In practice, each enclave would produce byte counts without timestamping them because their individual clock sources are untrusted (see Section~\ref{sec:stateless-filter-design}).
The byte counts are then collected in a timely manner and used for the optimization problem.}
A single enclave has the memory limit $M$ (e.g., 92 MB) and the bandwidth capacity $G$ (e.g., 10 Gb/s) as we have discussed in Section~\ref{sec:bottleneck-rule}. Then, we can decide the minimum number of enclaves as needed as $n_{min} = \lceil \max\left({\frac{1}{G}\sum_{i = 1}^{k}b_i ,\frac{ku}{M-v}}\right)\rceil$. To allow some space for optimization, the number of enclaves is taken as $n =\lceil \max\left({\frac{1}{G}\sum_{i = 1}^{k}b_i ,\frac{ku}{M-v}}\right) \times(1 + \lambda)\rceil$ where $\lambda \ge 0$ is an adjustable parameter for additional enclaves. 
%
We define real-valued variables $x_{i,j~(1 \leq i \leq k, 1 \leq j \leq n)}$ denoting the portion of bandwidth $b_i$ allocated to the $j$-th enclave, and binary variables $y_{i,j~(1 \leq i \leq k, 1 \leq j \leq n)}$ representing if rule $r_i$ is installed on the $j$-th enclave (i.e., $y_{i,j} = 1$). 
Based on the allocation plan indicated by $x_{i,j}$ and $y_{i,j}$, we consider $C_j = u \times \sum\limits_{i} y_{i,j} + v$ as the memory cost function, which is a linear function of the number of rules installed (where $u$, $v$ are constants).
Also, $I_j = \sum\limits_{i} x_{i,j}y_{i,j}$ is considered as the allocated bandwidth.
%
%
We present the detailed ILP formulation in Appendix~\ref{app:alp-formulation}.

\paragraph{Greedy algorithm to calculate filter rules}. 
Solving the above-mentioned optimization problem is inherently costly when $k \times n$ is large (e.g. $ > 10K$).
Thus, we propose a greedy algorithm (see Appendix~\ref{sec:greedy-algo}) that finds a sub-optional solution within a reasonably short time period.
The high-level intuition of the greedy algorithm is to pre-compute the two parameters---(1) the number of rules per enclave $h$ and (2) the bandwidth quota per enclave $g$---and arrange the rules and bandwidths for the obtained two parameters heuristically.  



\paragraph{Detecting misbehavior of untrusted components.}
The load balancer and other components outside of the enclave are untrusted and may misbehave.
For example, the load balancer may redirect to a filter the traffic flows that do not match with the filter rules assigned for that filter.
However, these misbehaviors can be easily detected by each filter by checking if it receives any packets that do {\em not} match the rules it receives from the master node.
Each enclave detects such load-balancing misbehavior, and reports it to the DDoS victim.
Note that if the load balancer drops the traffic flows that are supposed to be redirected to an enclave, it can be detected by the bypass detection of the auditable filters (see Section~\ref{sec:filter-design}).

%
%



\section{Implementation and Evaluation}
\label{sec:eval}

We implement a proof of concept of the \name filter using SGX and various optimizations (\S\ref{subsec:implementation}) and then evaluate its data-plane (\S\ref{sec:data-plane-performance}) and scalability performance (\S\ref{subsec:eval-scalablability}) for large attack volume and complexity.
%

\subsection{Implementation}
\label{subsec:implementation}
\begin{figure}[t]
	\centering
	\includegraphics[width=0.48\textwidth]{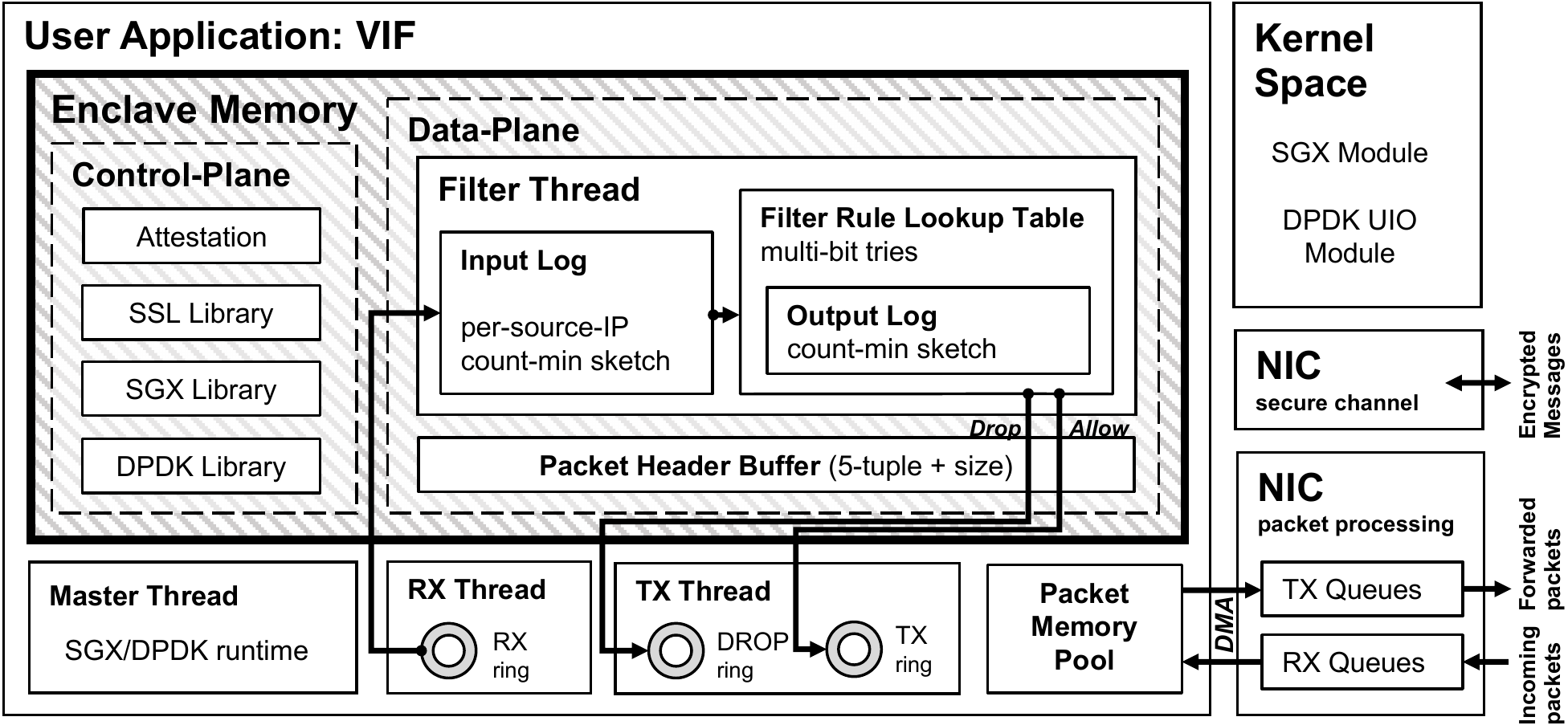}
	\caption{Implementation details of \name.}
	\label{fig:impl-overview}
\end{figure}

\paragraph{Overview}. 
We build the \name filter as a Linux userspace application with Intel SGX SDK 2.1 and DPDK 17.05.2 for high-speed packet processing. Figure~\ref{fig:impl-overview} shows the main components of the \name filter and the minimal trusted computing base (TCB) of code and data inside the enclave, which includes the entire control-plane and the key parts of the data-plane logic (e.g., packet logging and filtering). The control plane performs remote attestation and manages the keys for communication with a DDoS victim. The design of the data plane follows DPDK pipeline model, where three threads (i.e., RX thread, Filter thread, and TX thread) run on individual CPU cores and packets are passed between cores via DPDK lockless rings (i.e., RX ring, DROP ring, and TX ring). Every thread runs a small loop polling the hardware or software buffers in the previous stage, processes a batch of the packets, and passes it to the next stage in the pipeline. 

\begin{figure}[t]
	\centering
	\includegraphics[width=0.45\textwidth]{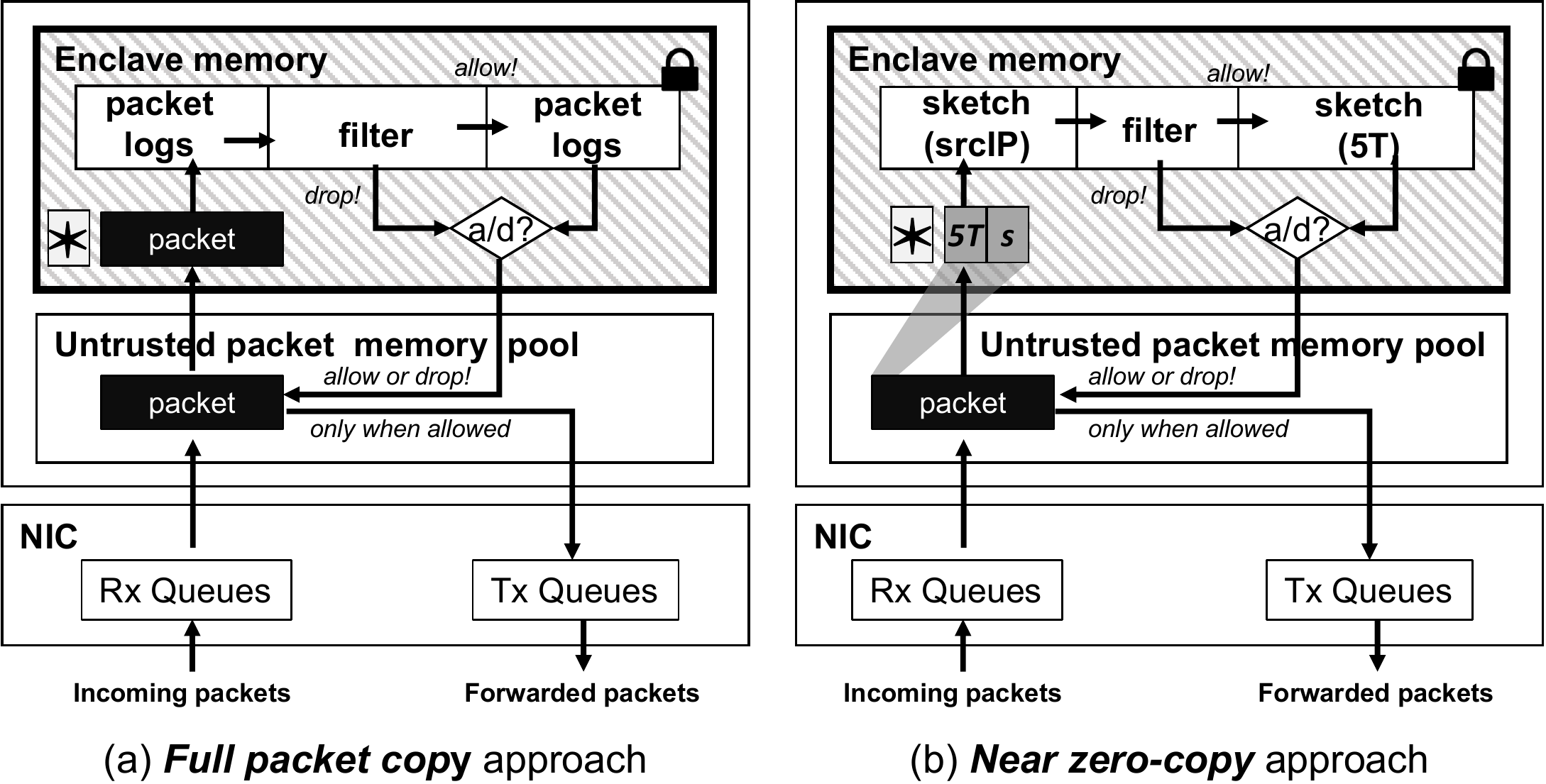}
	\vspace{-5pt}
	\caption{Two packet copy approaches for the auditable filter and packet logs. In the near zero-copy approach, we copy only the memory reference ($\ast$), the five tuple ($5T$), and the size ($s$) of the packet into the enclave.}
	\vspace{-10pt}
	\label{fig:near-zero-copy}
\end{figure}

\paragraph{Optimization: Near zero-copy design.}
%
For every incoming data-plane packet, a \name filter logs the packet, filters it based on the given filter rule set, and logs it again if it is allowed by the filter, as shown in Figure~\ref{fig:near-zero-copy}.
Figure~\ref{fig:near-zero-copy}(a) shows a naive approach, where a \name filter makes the entire copy of incoming packets into the enclave and operate these functions over the packet copies inside the enclave.
This {\em full-packet copy} approach can be considered as the baseline packet processing mechanism of other existing SGX-based middlebox applications (e.g., Tor nodes~\cite{kim2017enhancing,shinde2017panoply}, TLS middleboxes~\cite{naylor2017and,han2017sgx,goltzsche18endbox}, inter-domain routing~\cite{prixp,kim2015first}, and IDSs~\cite{trach2018shieldbox,shih2016s,goltzsche18endbox}), where secure operations over the full packet bytes are required (e.g., full packet read or encryption). 
However, this approach may incur too much overhead when performing line-rate processing due to the remaining EPC memory for a \name filter is only about 92 MB.

We thus minimize the dynamic memory usage and avoid the paging by copying only certain header fields into the enclave, which we call {\em near zero-copy} optimization. 
This allows more memory space for filter rules and the lookup table.
%
In particular, only a fraction of each packet's header fields (i.e., the five tuple fields, $5T$, and the packet size, $s$) are copied into the enclave memory along with the memory reference ($\ast$) of the packet, as shown in Figure~\ref{fig:near-zero-copy}(b).~\footnote{Such reduction of byte copies is allowed for our auditable filter applications but this does not necessarily apply to any other SGX-based middlebox applications.}
The copied data $\langle 5T,s \rangle$ represents the packet and is used for the logging functions and the filter operation. 
The memory reference $\ast$ is used to perform the corresponding operation (e.g., allow or drop) for the packet in the untrusted memory pool.

With the copied five tuple and the size, we first log each packet using a count-min sketch~\cite{cormode2005improved} (with $2$ independent linear hash functions, $64K$ sketch bins, and $64$ bit counters) for memory efficient (e.g., 1 MB) per-source-IP counters. 
The per-source-IP sketch for the incoming packets enables each neighboring ISPs of the filtering network to detect the `drop before filtering' bypass attack discussed in Section~\ref{sec:filter-bypass-detection}.
For forwarded packets, we also record another count-min sketch based on the full 5-tuple bits so that the victim network can detect bypass attempts.
The latency increased by the two sketch operations is negligible because only 4 linear hash function operations are conducted in the data-plane path.
Each counter has 64 bits and takes only around 1 MB EPC memory per instance of the count-min sketch.


\paragraph{Optimization: Reducing the number of context switches}. 
Another major overhead stems from the context switches when user application calls the enclave functions (ECall) or the enclaved function calls the outside functions (OCall). 
We address this performance degradation by reducing both types of calls in the filter thread: (1) \name only needs one ECall to launch the filter thread and initiates its polling; and (2) the filter thread makes no OCalls as the communication with other threads relies only on the software rings.

\paragraph{Trusted computing base (TCB).} Beyond the DPDK library containing about 64K source lines of code (SLoC), our \name filter contributes to the TCB only 1,206 SLoC which includes the modification of DPDK \texttt{ip\_pipeline} (1044 SLoC) and the packet logging and near zero-copy functions (162 SLoC).

\subsection{Line-rate Data-plane Performance}
\label{sec:data-plane-performance}
\paragraph{Testbed Setup}. We test our implementation with two machines: one is a packet generator and one deploys \name filter. 
The packet generator has an Intel E5-2630 v3 CPU (2.40 GHz, 8 cores) and 32 GB memory. 
The filtering machine has an Intel i7-6700 CPU (3.40 GHz, 4 cores) and 8 GB memory. 
Both have 10 GbE Intel X540-AT2 network cards and run Ubuntu 16.04.3 LTS with Linux kernel 4.10. 
On the packet generator machine, we use \texttt{pktgen-dpdk} 3.4.2 to generate the traffic saturating the 10 Gb/s link between the two machines.



\begin{figure}[t!]
	\centering
	\includegraphics[width=0.45\textwidth]{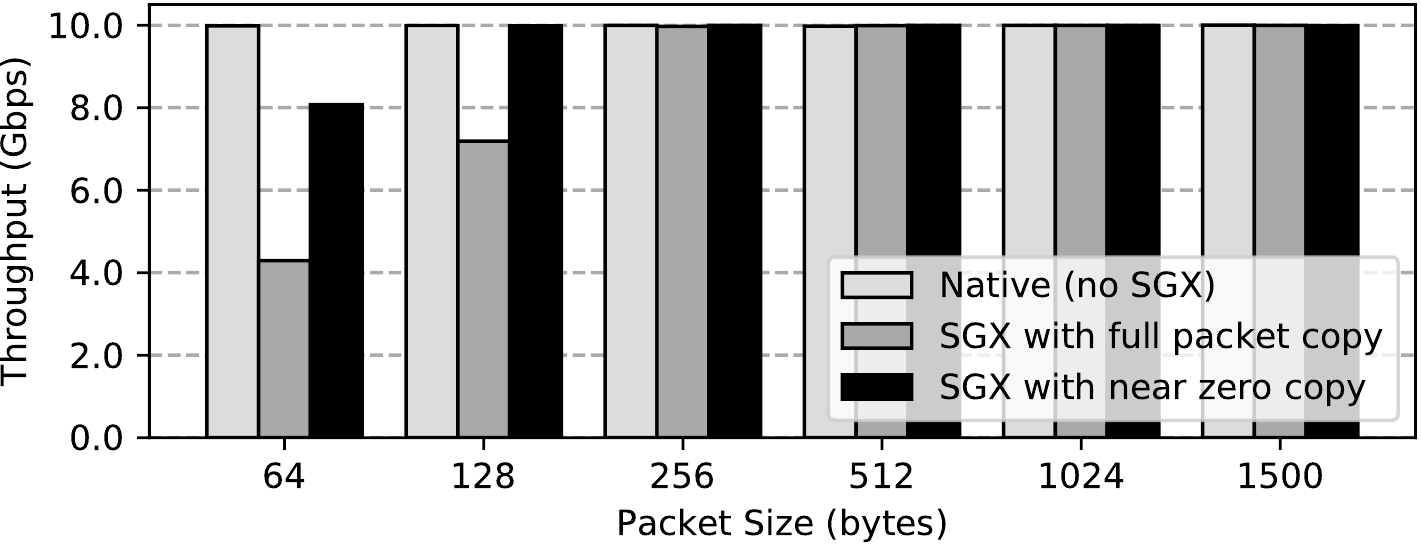}
	\caption{Throughput performance in bit-per-second for varying packet sizes and 3,000 rules with three implementation versions: (1) Native (no SGX), (2) SGX with full packet copy, and (3) SGX with near zero copy.}
	\vspace{-15pt}
	\label{fig:throughput-bps}
\end{figure}

\paragraph{Throughput Performance.}
We benchmark the maximum throughput performance of the filter with the packet size of 64, 128, 256, 512, 1024, and 1500 bytes for three different versions of the \name filter implementations:
(1) native filter without SGX, 
(2) SGX-based filter with full packet copy, and
(3) SGX-based filter with near zero-copy.


Figure~\ref{fig:throughput-bps} shows the throughput performance for varying packet sizes for the three implementations. 
For the packet sizes of 256 Byte or larger, all the three implementations achieve the full line-rate of 10 Gb/s. 
With small packet sizes, however, we observe some degradation due to the use of SGX.
Particularly, when we make full packet copies for each incoming packet, the filter experiences significant throughput degradation. 
The near zero-copy implementation demonstrates {8 Gb/s throughput performance even with 64 Byte packets and 3,000 filter rules}.
Additionally, we present the experiment results of \name evaluation in packet per second metric in Appendix~\ref{sec:throughput-performance}.

\paragraph{Latency Performance.} 
We also measured the latency for the near zero-copy version with various packet size starting from 128 bytes. The results are 34 $\mu$s (128 bytes), 38 $\mu$s (256 bytes), 52 $\mu$s (512 bytes), 80 $\mu$s (1024 bytes), 107 $\mu$s (1500 bytes). All the measurements are average latency over 10-second run with 8 Gb/s constant traffic load, which are reported by \texttt{pktgen}'s latency measurement function.

\paragraph{Connection-Preserving Filtering Performance.}
We evaluate the detailed performance of connection-preserving filtering.
We present the result in Appendix~\ref{sec:connection-performance}.

\paragraph{Remote attestation performance.}
Our detailed remote attestation performance can be found in Appendix~\ref{app:remote-attestation-performance}.


\subsection{Scalable Filter Rule Distribution}
\label{subsec:eval-scalablability}

We evaluate the solving performance of the mixed ILP optimization problem described in Section~\ref{sec:optimization-multiple-enclaves} with the CPLEX solver in a server-grade machine with 20 cores. 
We use 3,000 or more filter rules that would cause the throughput degradation of each \name filter. 
In this evaluation, we consider that the total traffic rate going through the entire VIF filter is 100 Gb/s. 
The incoming traffic distribution across the filter rules follows a lognormal distribution. 

With the number of rules more than 3,000 and the number of enclaves more than 10, we find that the CPLEX's mixed ILP solver cannot return the optimal solutions within any reasonable time period. 
To evaluate the effectiveness of our greedy algorithm in Section~\ref{sec:optimization-multiple-enclaves}, that is, how close it is to the optimal solutions, we use a small number of filter rules ($10\le k \le 15$) and confirm that the difference between the optimal cost function calculated by the CPLEX's mixed ILP solver and the results from our greedy algorithm is only 5.2\%.

\begin{table}[t]
	\centering
	\footnotesize
	\caption{Execution times for the ILP solution and the greedy algorithm solution. The CPLEX's mixed ILP solver is configured to stop when found sub-optimal solutions.}
	\label{tab:exe-time-for-ILP-greedy}
	\begin{tabular}{r r r}
		\toprule
		Number of rules ($k$) &  CPLEX (sub-optimal) & Greedy  \\
		\midrule
		5,000  &  210.49s & 0.31s  \\
		10,000 &  772.43s & 0.50s  \\
		15,000 & 1,614.96s & 0.73s  \\
		\bottomrule
	\end{tabular}
\end{table}

We now compare the execution time of the CPLEX's mixed ILP solver and our greedy algorithm when the number of rules is between 5,000 and 15,000 and show the results in Table~\ref{tab:exe-time-for-ILP-greedy}.
To measure the execution time of the CPLEX's mixed ILP solver for our optimization problem, we configure the solver to stop earlier when it finds a first, sub-optimal solution. 
As shown in Table~\ref{tab:exe-time-for-ILP-greedy}, the CPLEX solver even requires about 200 -- 1,600 seconds to find the sub-optimal solutions, which are unacceptably slow for the dynamic filter rules redistribution operations.
On the other hand, our greedy algorithm runs {three orders of magnitude faster than the CPLEX solver} with the same number of filter rules.

\begin{figure}[t!]
	\centering
	\includegraphics[width=0.45\textwidth]{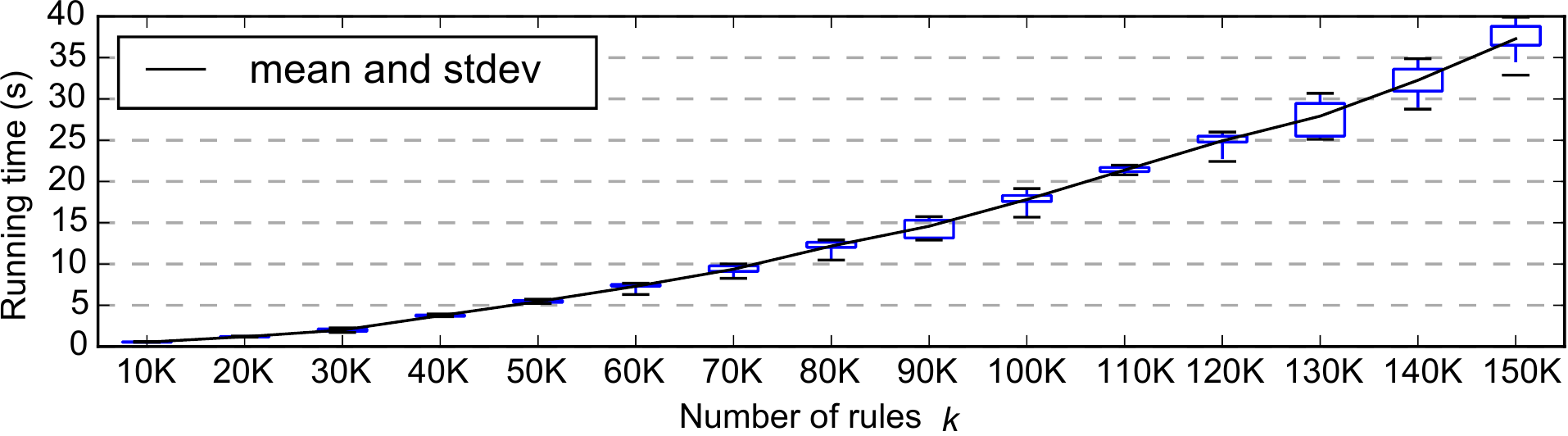}
	\caption{Average time taken (in second) to complete the heuristic algorithm for optimizing filter rules across multiple enclaves for varying number of rules $k$.}
	\label{fig:greedy-algo-time}
\end{figure}

Figure~\ref{fig:greedy-algo-time} shows the extended experiments on the execution times of the greedy algorithm for varying number of rules at a much larger range. 
We also run this experiment with the total traffic bandwidth of 500 Gb/s, which follows the log-normal distribution. 
In all the range we test (10K--150K filter rules), the greedy algorithm requires no more than 40 seconds.
This enables a near real-time dynamic filter rule re-distribution for large numbers of \name filters. 

\section{Practical Deployment at IXP}
\label{sec:deployment-proposals}

\name has a generic architecture designed for any transit networks (e.g., Tier-1 or large Tier-2 ISPs); yet, as the first deployment model, we suggest to deploy it in major Internet exchange points (IXPs).
In this section, we present why IXPs, among other transit infrastructure, are the ideal places to introduce our verifiable in-network filtering (\S\ref{sec:ixp-overview}). 
We also provide a deployment example of \name at an IXP (\S\ref{subsec:ixp-deployment}) and then evaluate the effectiveness of \codename at IXPs against DDoS attacks with two real attack source data (\S\ref{subsec:against-ddos-exp}).
Finally, we provide a simple cost analysis for deploying \name service at an IXP for filtering up to 500 Gb/s traffic (\S\ref{subsec:deployment-cost}).



\subsection{Internet Exchange Points (IXPs)} 
\label{sec:ixp-overview}


IXPs are physical locations where multiple autonomous systems (ASes) peer and exchange traffic and BGP routes. 
Essentially, an IXP is a large layer-2 fabric and connects ASes (e.g., ISPs, content providers, cloud providers) in close proximity. 
IXPs provide great convenience to ASes in making peering relationship with many (e.g., hundreds or thousands) other ISPs without the hassle of individual bilateral agreements. 
The Internet currently has more than 600 globally distributed IXPs~\cite{caida-ixp} and some large IXPs serve multi-Tera b/s traffic volume, which is comparable to large Tier-1 ISPs~\cite{ager2012anatomy,chatzis2013there}.


Recently, as more video content providers and large cloud providers rely on IXPs for lower cost but faster transit of their traffic, emerging {\em value-added services} are expected from IXPs; e.g., application-specific peering,  inbound traffic engineering,  wide-area server load balancing, and redirection through middleboxes~\cite{gupta2015sdx}.
New innovation for these value-added services has been possible because IXPs have a flexible architecture model (especially, compared to traditional transit networks, such as ISPs).
An IXP usually consists of a single data center in a single physical facility;
thus, software-based architecture available for data centers can be easily adopted; see software-defined IXPs~\cite{gupta2015sdx, gupta2016authorizing,gupta2016industrial}.


Due to their topological centrality, however, IXPs unfortunately often suffer from collateral damage when other networks are targeted by DDoS attacks~\cite{pospichal2017new}. 
Worse yet, IXPs are sometimes directly targeted by DDoS attacks; see an attack incident in 2016 against multiple IXPs: AMS, LINX, and DE-CIX~\cite{bright2013can}. 
A traffic filtering service could easily be a natural next value-added service for IXPs~\cite{dietzel2016blackholing}.

\begin{figure}[t!]
	\centering
	\includegraphics[width=0.45\textwidth]{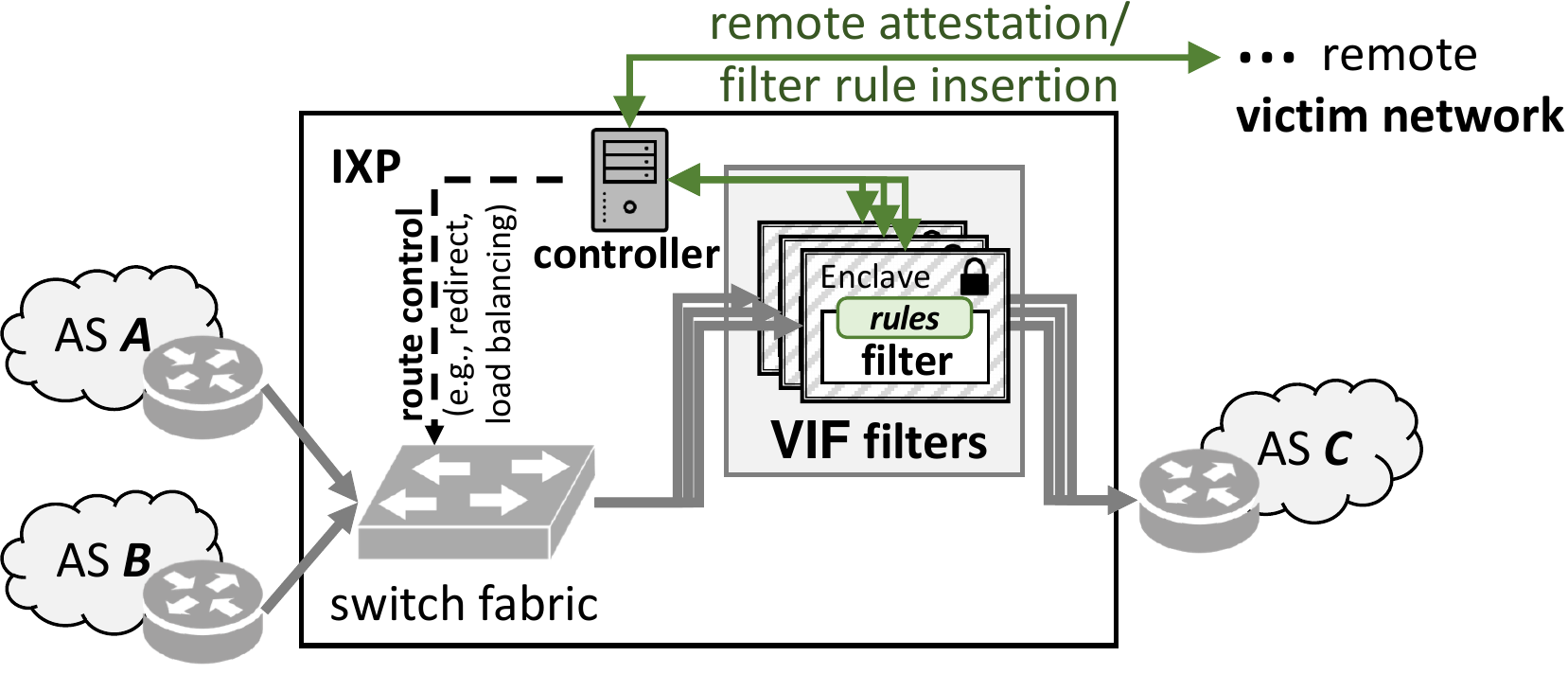}
	\caption{Deployment example of \name at IXP.}
	\label{fig:ixp-deployment}
\end{figure}

\subsection{Deployment example at IXP}
\label{subsec:ixp-deployment}

We consider the \name IXP has a generic architecture that includes a layer-2 switching fabric, a route server (which is not highlighted in our paper), and a logical central controller for software-defined switches~\cite{gupta2015sdx,gupta2016industrial}.
Figure~\ref{fig:ixp-deployment} illustrates a deployment example of \name at an IXP.
The filtering IXP sets up one or more commodity servers with SGX support. 
%
When a victim network is under DDoS attack, it contacts the controller of the \name IXP via an out-of-band channel.\footnote{We assume this channel is available even when the victim network is under attacks.
ISPs traditionally have maintained out-of-band channels (e.g., external email servers, telephone lines~\cite{dots-slides}) for inter-ISP communication.} 
As suggested in~\cite{ramanathan2018senss}, the victim network can easily authenticate to the IXP via Resource Public Key Infrastructure (RPKI)~\cite{lepinski2012infrastructure}.
The victim network asks the filtering IXP to create one or more SGX filters and audits it by receiving the validation attestation report(s). 
After being convinced that the filters have been set up properly (i.e., the remote attestation is done), the victim network establishes a secure channel with the enclaves (e.g., TLS channels) and submits the filtering rules. 
The load balancing algorithm at the IXP controller receives the rules from the filter parallelization (see Section~\ref{sec:optimization-multiple-enclaves}) and accordingly controls the switches to distribute traffic destined to the victim network to the enclaved filters.
The \name IXP eventually learns and analyzes all the rules in this step.
Finally, the enclave filters perform packet filtering based on the submitted rules and forward the allowed traffic to the victim network.






\subsection{Effectiveness of \codename at IXPs against DDoS Attacks}
\label{subsec:against-ddos-exp}

\begin{figure}[t!]
    \centering
    \begin{subfigure}[t]{0.24\textwidth}
        \centering
        \includegraphics[width=\textwidth]{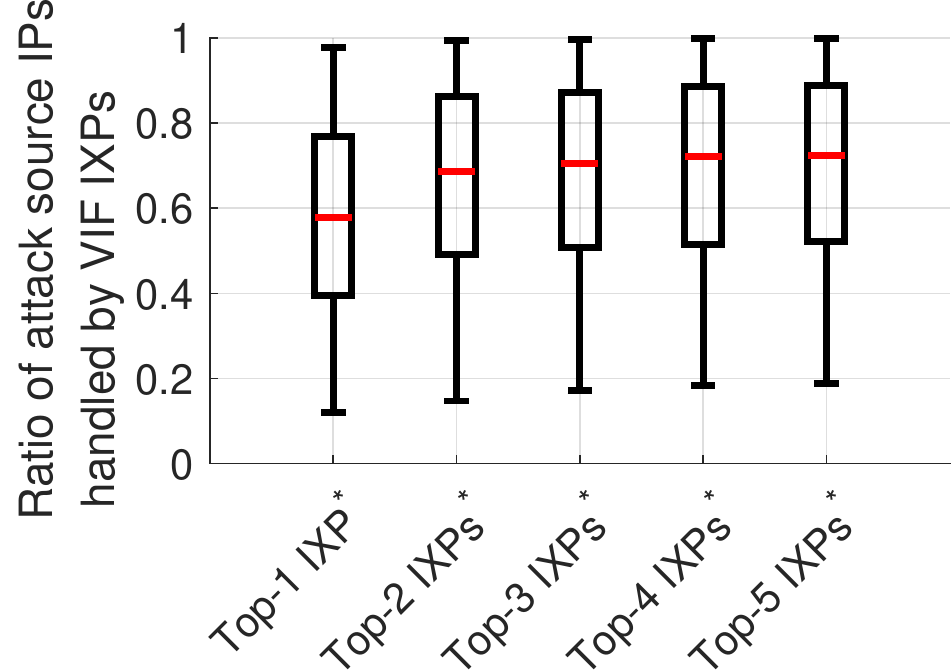}
        \vspace{-10pt}
        \caption{Vulnerable DNS resolvers}
        \vspace{-5pt}
        \label{fig:ixp-openres}
    \end{subfigure}
    \begin{subfigure}[t]{0.22\textwidth}
        \centering
        \includegraphics[width=\textwidth]{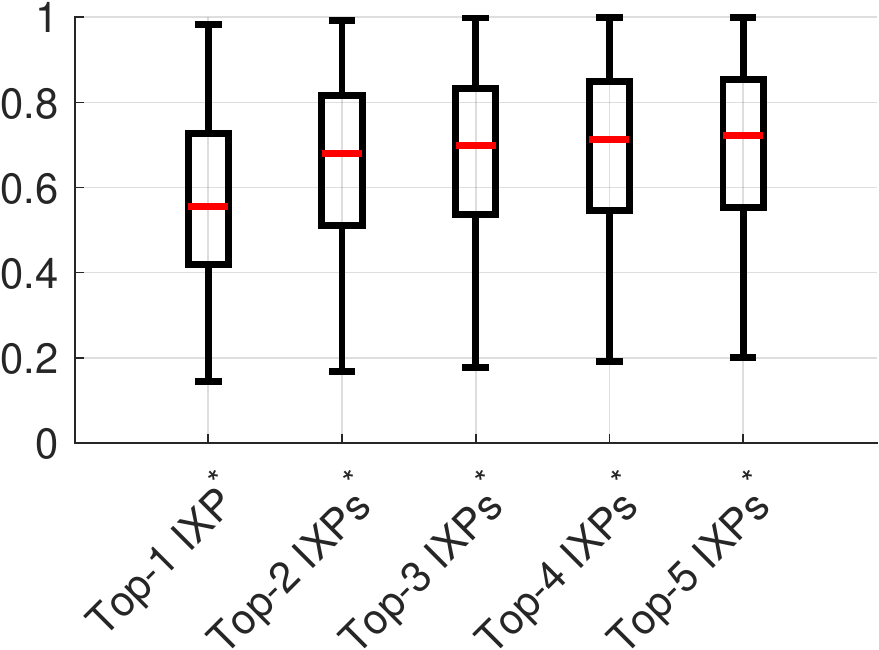}
        \vspace{-10pt}
        \caption{Mirai botnets}
        \vspace{-5pt}
        \label{fig:ixp-mirai}
    \end{subfigure}
	\vspace{-5pt}
    \caption{The ratio of attack sources that are handled by the \name filters for the two attack source data. $(^*)$: Top-$n$ IXPs denote the $n$ largest IXPs in each of the five regions, see Table~\ref{tab:top-ixps} in {{Appendix~\ref{sec:top-ixp}}}.}
    \vspace{-7pt}
    \label{fig:ixp-coverage}
\end{figure}

%
We analyze how much DDoS attack traffic can be filtered by \name at IXPs with two real attack source data: 3 million vulnerable open DNS resolver IP addresses~\cite{wessels2014open} and 250 thousand Mirai bot IP addresses~\cite{mirai}.

\paragraph{Simulation setup.}
In our inter-domain routing simulation, we use the CAIDA Internet measurement data with the inferred AS business relationship~\cite{caida} and the peering membership of world-wide IXPs~\cite{caida-ixp}.
We randomly choose 1,000 Tier-3 ISPs as the DDoS victims and consider that each victim receives attack traffic from all the attack sources (e.g., open resolvers and bots) in each case.
To determine a traffic forwarding path between autonomous systems (ASes), we assume that each of them applies the following widely adopted BGP routing policies in order~\cite{gill2013survey,gao2001inferring}:
(1) the AS prefers customer links over peer links and peer links over provider links;  (2) the AS prefers the shortest AS-path length route; and
(3) if multiple best paths exist, the AS uses the AS numbers to break the tie. 



We assume that the victim network establishes \name sessions with several largest IXPs (e.g., the biggest IXPs in each of the five regions, as shown in Table~\ref{tab:top-ixps}, see {{Appendix~\ref{sec:top-ixp}}}). 
We compute the ratio of flows from the attack IP addresses to the victim network that are handled by at least one of the established \name filters at the selected IXPs.
A traffic flow is said to be transited at an IXP if it traverses along an AS-path that include two consecutive ASes that are the members of the IXP.

\paragraph{Results.}
Figure~\ref{fig:ixp-coverage} shows how many attack flows can be effectively handled by the in-network filters if installed in some large IXPs.
The box-and-whisker plots show the distribution of the ratio of handled attack IPs when Top 1--5 biggest IXPs in the five regions (thus, 5--25 IXPs globally in total) perform in-network filtering service for DDoS defense. 
In each plot, the solid lines represent the first and the fourth quartile of the data set and the ends of the whisker indicate the 5th- and 95th-percentiles. 
Also, the red band inside the box represents the median.

Even when a single IXP in each region (thus, total five IXPs worldwide) adopts the \name filters, the majority of both attack sources (e.g., vulnerable resolvers, botnet) are handled by the \name IXPs.
Approximately, 60\% of attack mitigation is expected for the median cases, and 70-80\% mitigation can be achieved for the top quarter cases.
As more IXPs adopt the \name filters, even more effective mitigation is achieved.
Particularly, Top-5 IXPs per these regions appear to be sufficient enough to offer more than 75\% attack mitigation for the median cases, and 80-90\% of attack mitigation for the upper quarter cases. 


\subsection{Deployment Cost Analysis}
\label{subsec:deployment-cost}
Let us provide a ballpark estimate of the cost of deploying \name at an IXP to handle 500 Gb/s of traffic.
Note that the 500 Gb/s filter capacity at a {\em single} IXP appears to be sufficient because the attack volume at each IXP can be much lower than the aggregated volume at the victim network.
For instance, it would require only a few \name IXPs with similar capabilities to mitigate the biggest DDoS attack ever recorded with 1.7 Tb/s attack traffic~\cite{zdnet2018ddos}.

Our experiment in Section~\ref{sec:data-plane-performance} shows that a near full line-rate performance of 10 Gb/s per server with four SGX cores is easily achieved. 
Thus, to handle 500 Gb/s attack traffic, an IXP needs to invest in 50 modest SGX-supporting commodity servers, which would require only one or two server racks.
With a commodity server cost is approximately US\$ 2,000, the filtering IXP only needs to spend for one-time investment for US\$ 100K to offer an extremely large defense capability of 500 Gb/s.
The capital expenditure can be borne by the member ASes (hundreds or thousands) and/or can be amortized by the service fees if the filtering service is economically compensated by the payment from the victims~\cite{ramanathan2018senss}.
A rigorous economic analysis of \name operations in IXPs is out of the scope of this paper and is left for future work.

\section{Discussion}
\label{sec:discussion}

We discuss several commonly asked questions regarding the designs and usages of the \name system. 

\paragraph{What if victim networks cause denial-of-service by blocking arbitrary packets?}
Malicious victim networks cannot exploit \name and launch new DoS attacks because filter rules are first validated with RPKI. 

\paragraph{Can victim networks exploit \name?}
Even when it is {\em not} under attack, an {\em alleged} DDoS-victim network can make use of \name to filter some flows and reduce its operating cost. 
Regardless of the underlying motivation of \name requests, however, we do not see this as a serious threat to the \name system model.
First, a \name filtering network may also save its operating cost due to the reduced upstream bandwidth usage. 
If not, a \name filtering network can always refuse to serve any \name request. 
Second, a victim network would request filtering of flows that otherwise would be filtered by itself anyway. 
Third, if payment channels exist between the two networks~\cite{ramanathan2018senss}, this is not considered as the abuse of the service but a legitimate outsourced filtering.

\paragraph{Incrementally deployable?}
Incremental deployability of new services is a perennial problem in a global Internet. 
IXPs have been recently highlighted as the practical Internet infrastructure where emerging services can be deployed and used incrementally~\cite{gupta2015sdx,gupta2016industrial}.
A \name instance installed at even single IXP can readily help reduce DDoS traffic by the direct victims with strong verifiability property. 

%

\paragraph{Can \name handle any DDoS attacks?}
We do {\em not} claim that in-network filtering can handle all types of DDoS attacks.
\name demonstrates that it can effectively handle volumetric attacks.
However, \name may not be as equally effective against non-volumetric attacks (e.g., application-layer attacks), and thus we suggest that \name is used as an additional layer of defense mechanism, which is orthogonal to existing cloud-based defense solutions. 


\paragraph{What does a DDoS victim do when misbehavior is detected?}
In \name, we suggest two networks create a filtering contract only between them. 
Therefore, any one of them can abort the temporary contract whenever one party detects misbehavior. 
\vspace{-5pt}
\section{Related Work}
\label{sec:rw}

Network DDoS attacks and defenses have been extensively studied in the last 2--3 decades~\cite{mirkovic2004taxonomy}. 
Here, we summarize a few categories of DDoS defenses and related projects.

\vspace{-5pt}
\subsection{In-network Filtering}

The idea in-network filtering has been the core idea of many DDoS mitigation proposals.  
There are two prominent approaches to implementation: dynamic filtering and capability-based approaches.
Dynamic filtering suggests that the destination ISP requests the ISPs on the forwarding paths to install filter rules at the time of attack, for instance as proposed in
Pushback~\cite{mahajan2002controlling}, D-WARD~\cite{mirkovic2002attacking}, AITF~\cite{argyraki2005active}, StopIt~\cite{liu2008filter}.
Capability-based approaches embed capabilities in the packet flows, which can be controlled by the destination hosts to authorize flows in upstream, as proposed in SIFF~\cite{yaar2004siff}, TVA~\cite{yang2005limiting}, and Portcullis~\cite{parno2007portcullis}. 

Closest to our work is the recent SENSS defense architecture by Ramanathan et al.~\cite{ramanathan2018senss}.  
SENSS proposes to install DDoS-victim submitted filters at a small number of major ISPs. 
Ramanathan et al. show that in-network filtering at only four major ISPs in the US would have stopped the Dyn attack happened in 2016~\cite{york2016dyn}.
SENSS also suggests an automated payment channel between a DDoS victim and a filtering ISP so that the ISP can get compensation for the extra filtering tasks. 
Although the idea is solid and evaluation is promising, the SENSS proposal lacks the filtering verifiability and thus allows several undetectable misbehaviors of the filtering ISPs.\footnote{Ramanathan et al.~\cite{ramanathan2018senss} sketch a reputation-based mitigation, which can be used together with our \name proposal.}



Unlike previous in-line filtering proposals, our system \name focuses on the highly desired but yet-unaddressed security property for in-network filtering system, i.e., the verifiable filter, and demonstrates its feasibility and scalability.

\vspace{-5pt}
\subsection{Network Function Virtualization with Trusted Hardware}
We categorize some of them:
\begin{packeditemize}
	\item {\bf Middleboxes:} Various network middleboxes have been tested with the SGX capability. 
	TLS middleboxes~\cite{naylor2017and,han2017sgx} demonstrate that an SGX-protected middlebox can handle thousands of TLS sessions without violating their end-to-end encryption. 
	ShieldBox~\cite{trach2018shieldbox} and Trusted Click~\cite{trusted-click} demonstrate that SGX can protect the Click modular router to implement various trusted network functions.  
	S-NFV~\cite{shih2016s} also discusses general policy, data privacy issues of network functions. 
	LightBox~\cite{lightbox} demonstrates the line-rate performance for simple secure packet processing functions. 
	Snort-SGX~\cite{kuvaiskii2018snort} also demonstrates line-rate performance of Snort 3 along with a DPDK network layer.
	SafeBricks~\cite{poddar2018safebricks} implements a highly modularized and safe network function architecture in the Intel SGX platform.
	
	Our main contribution is not merely an integration of a simple flow filter function and an SGX architecture but more on addressing scalability and filter-rule violations at network layer that are specific to verifiable in-network filtering defense systems. 
	
	\item {\bf Privacy-preserving systems:} 
	Several systems demonstrate that SGX-based network functions can improve the privacy of anonymity systems: SGX-protected Tor nodes~\cite{kim2017enhancing,shinde2017panoply},
	\item {\bf Inter-domain routing:}  
	Also, several secure inter-domain routing applications have been proposed to exploit the security guarantees of Intel SGX~\cite{prixp,kim2015first}.
	\item {\bf Verifiable accounting:} 
	There also have been some prototype systems that enable the outsource network functions to securely measure the amount of resources used for the requested tasks (e.g., \cite{tople2018vericount} in an SGX platform, \cite{vee13} in a TPM platform).
\end{packeditemize}

With \name we investigate a unique design point of auditable traffic filters. 
Particularly, our contribution of \name is in the design of auditable filters that can handle ever-growing large-scale attack volume and complexity with the SGX support.
%
%
%
%
%
%
%

\vspace{-5pt}
\subsection{Cloud-based DDoS Mitigations}

The predominant DDoS defense in practice today is an overlay-based filtering approach, such as cloud-based scrubbing services, that performs outsourced filtering in a third-party network on behalf of the DDoS victims (e.g., AWS-Shield~\cite{aws-shield}, Radware DefensePro~\cite{radware}).
Overlay-based filtering approaches are popular particularly because they require {\em no} changes to the current Internet
architecture. Recent works have proposed advances in such overlay
filtering using middleboxes~\cite{liu2016middlepolice,gilad2016cdn} and proposals
have discussed large-scale filtering locally at
ISPs~\cite{fayaz2015bohatei}.
However, end users are not satisfied with the status quo. Reports
suggest that relying on third-party providers centralizes the DDoS
marketplace~\cite{gilad2016cdn}; costs for small and medium-size victims
are high and services are left to the discretion of large service
providers~\cite{guardian2016ddos,gilad2016cdn}.


Unlike these proposals, \name does not rely on the cloud or the local victim network's capability but directly establishes filtering rules at the transit networks. 

%
%

%
%
%

\vspace{-5pt}
\section{Conclusion}
In-network filtering has numerous known advantages over other proposed DDoS defenses, but the deployment of any of its variants has been stifled. 
Our proof of concept \name system addresses one of the core, but largely neglected, problems of source filtering---lack of filtering verifiability---and demonstrates that verifiable in-network filtering is indeed possible with a practical hardware root of trust support. 
We hope that our study renews discussion on the deployment of in-network filtering in the IXPs and encourages more sophisticated yet auditable filter designs, such as stateful firewalls.


\bibliographystyle{IEEEtranS}
\bibliography{paper}
\appendices

\section{Flow-aware Filter Design}
\label{sec:filter-design-points}

We consider deterministic and non-deterministic filter rule types:
\begin{packeditemize}
	\item A {\em deterministic filter rule} defines a static \{{\tt ALLOW, DROP}\} decision for a specified flow; and
	\item A {\em non-deterministic filter rule} expresses only the static probability distribution ($P_{\tt ALLOW}$, $P_{\tt DROP}$, where $P_{\tt ALLOW}+P_{\tt DROP}=1$) for a specified flow and the final filter decision for each exact flows is made by the \name filter.
	We guarantee the connection-preserving property in \name filters so that all the packets in a TCP/UDP flow are allowed or dropped together.
\end{packeditemize}

The execution of non-deterministic rules with the {\em connection-preserving property} can be implemented in two different ways:
\begin{packeditemize}
	\item {\bf Hash-based filtering.} For each incoming packet $p$, we compute the cryptographic hash (e.g., SHA-256) of its five-tuple bits and the enclave's secrecy to make filtering decision based on the given probability distribution. For example, packet $p$ is allowed if $H(\texttt{five-tuple-bits}||secrecy) < (2^{256}-1) \times p_{\texttt{ALLOW}}$ with $H(\cdot)=$SHA-256; and
	\item {\bf Exact-match rule filtering.} For each TCP/UDP connection, the filter installs an exact-match rule with a filtering decision randomly chosen based on the given probability distribution. 
\end{packeditemize}
Note that the two design points have different advantages and disadvantages. 
The hash-based filtering has a smaller memory footprint for lookup table but it incurs per-packet additional latency for cryptographic hash operations. 
In contrast, the exact-match filter design tends to have shorter per-packet processing time since it executes only one lookup but it requires a larger memory footprint for lookup tables and adds latency for frequent lookup table updates. 
We propose a {\em hybrid} design where hash-based filtering is performed for new flows until these new flows are installed with exact-match rules at every rule update period (e.g., 5-40 seconds).

\section{Dynamic Test of Intermediate ASes}
\label{app:dynamic-test-intermediate-ases}
A victim network finds and avoids suspicious ASes that drop the \name-allowed packets. 
Particularly, we utilize the well-known BGP {\em poisoning-based} inbound rerouting techniques (e.g., LIFEGUARD~\cite{katz2012lifeguard} and Nyx~\cite{smith2018routing}) to reroute (or detour) inbound traffic and avoid traversing any intermediate ASes for a short period of time (e.g., a few tens of seconds). 
These BGP poisoning-based rerouting technologies do {\em not} require any inter-AS coordination and thus the victim network can independently test if any intermediate ASes drop packets (even without the \name IXP's agreement).

If misbehavior of an intermediate AS is detected by the victim network, then the misbehaving AS can be avoided for the extended period of time (at least during the \name session) for auditable filtering.
Or, if the victim network continuously witnesses that \name-allowed packets are dropped continuously when dynamically changing the inbound routes, it may conclude that the \name IXP itself has been misbehaving.
The victim network can then discontinue the \name contract with the \name IXP at its discretion.

Note that we do not consider extremely adverse network adversaries, such as dropping all the packets between the \name IXP and the victim network, which cannot be handled properly by any possible defenses in the current Internet architecture. 

\section{ILP Formulation for Multi-Enclave Optimization}
\label{app:alp-formulation}

Our goal is to fully utilize the available resource on $n$ enclaves in terms of the bandwidth and memory, without triggering the performance degradation. Also, the load on each enclave should be balanced, in order to reduce the chance of any enclave getting closer to the limit of $G$ or $M$. Hence, the maximum $C_j$ and maximum $I_j$ should all be as small as possible, as shown in Equation~\ref{eq:c1}. Note that a constant coefficient $\alpha$ is used to balance two maximums in the sum.
Because of the capacity limit of a single enclaved filter, any filter should have less than $M$ memory consumption (Equation~\ref{eq:c2}) and less than $G$ bandwidth load (Equation~\ref{eq:c3}).
Since $b_i$ is distributed to multiple filters, so their allocated bandwidth with respect to rule $i$ should sum up to the value of $b_i$ (Equation~\ref{eq:c4}).
Two decision variables are not independent since when $y_{i,j} = 0$, $x_{i,j}$ should never be a positive value. (Equation~\ref{eq:c5}).

\begin{figure}[t]
   \begin{mdframed}
   	\footnotesize
       \raggedright
       $$\textrm{Minimize}\quad z$$
       \begin{align}
        s.t.~~~~~~~~~&\forall p,q: z \geq \alpha \big(u \sum_{i = 1}^{k} y_{i,p} + v \big) + \sum_{i = 1}^{k} x_{i,q}y_{i, q} \label{eq:c1}\\
        &\forall i: u \cdot \sum_{j = 1}^{n} y_{i,j} + v \leq M \label{eq:c2}\\
        &\forall j: \sum_{i = 1}^{k} x_{i,j} y_{i,j} \leq G \label{eq:c3}\\
        &\forall i: \sum_{j = 1}^{n} x_{i,j} = b_i \label{eq:c4}\\
        &\forall x_{i,j}, y_{i,j}: (1 - y_{i,j}) x_{i,j} = 0 \label{eq:c5}\\
        &\forall x_{i,j} \geq 0 \quad \textrm{and} \quad \forall y_{i,j} \in \{0, 1\} \label{eq:c6}
       \end{align}
   \end{mdframed}
	\vspace{-5pt}
   \caption{ILP formulation for the optimal rule distribution. }
   \vspace{-15pt}
   \label{fig:ilp-formal}
\end{figure}

\section{Greedy Algorithm for Scalable Filter Design}
\label{sec:greedy-algo}

The pseudocode in Algorithm~\ref{algo:fair-comp} summarizes the greedy algorithm we use for the filter rule distribution problem for the scalable \name filter design in Section~\ref{sec:optimization-multiple-enclaves}.

\alglanguage{pseudocode}
\begin{algorithm}[ht]
	\caption{Greedy algorithm for rule distribution and bandwidth allocation}
	\label{algo:fair-comp}
	\begin{algorithmic}[1]
		\scriptsize
		\Procedure{GreedySolver}{$b_1, b_2, ...b_k$, $M$, $G$, $\lambda$, $u$, $v$} 
		\State $B \gets \{b_1, b_2, ..., b_k\}$, $g \gets \frac{1}{n}\sum_{i=1}^{k} b_i$, $h \gets \frac{k}{n}$
		
		\While{$g \leq G $ \textbf{and} $h \leq (M - v)/u $}
		\State $X \gets$\textsc{AssignBandwidth}($B$, $h$, $g$, $n$)
		\If{$X \neq \emptyset$}
		\State \Return $X$
		\EndIf
		\State $g \gets g + \Delta g$
		\If{$g > G$}
			\State $h \gets h + \Delta h$, $g \gets \frac{1}{n}\sum_{i=1}^{k} b_i$
		\EndIf
		\EndWhile
		\EndProcedure
		
		\Procedure{AssignBandwidth}{$B$, $h$, $g$, $n$}
		\State $X \gets \emptyset$, $j \gets 1$ 
		\While{$B \neq \emptyset $ \textbf{and} $ j \leq n$}
		\State $r \gets g$, $c \gets 0$ \Comment{$c$: remaining bandwidth for filter $j$; $f$: rule counter}
		\While{$B \neq \emptyset $ \textbf{and} $c \leq h$}
		\State $b_{i} \gets \textrm{PopMin}(B)$
		\If{$b_{i} < r $ \textbf{and} $ j + 1 \leq h$}
		\State $x_{i,j} \gets b_{i}$, $X \gets X \cup \{ x_{i,j}\}$, $c \gets c + 1$, $r \gets r - b_{i}$
		\State \textbf{continue}
		\EndIf
		\State $B \gets B \cup \{b_{i}\}$, $b_{i} \gets \textrm{PopMax}(B)$
		\If{$b_{i} \leq r$}
		\State $x_{i,j} \gets b_{i}$, $X \gets X \cup \{ x_{i,j}\}$, $c \gets c + 1$, $j \gets j + 1$
		\Else
		\State $x_{i,j} \gets r$, $X \gets X \cup \{ x_{i,j}\}$, $b_{i} \gets b_{i} - r$, $B \gets B \cup \{b_{i}\}$
		\EndIf
		\State \textbf{break}
		\EndWhile
		\EndWhile
		\If{$B = \emptyset$} \State \Return $\emptyset$ \EndIf
		\State \Return $X$
		\EndProcedure
	\end{algorithmic}
\end{algorithm}

\section{Additional Evaluation of Throughput Performance of \name}
\label{sec:throughput-performance}


\begin{figure}[t!]
	\centering
	\includegraphics[width=0.45\textwidth]{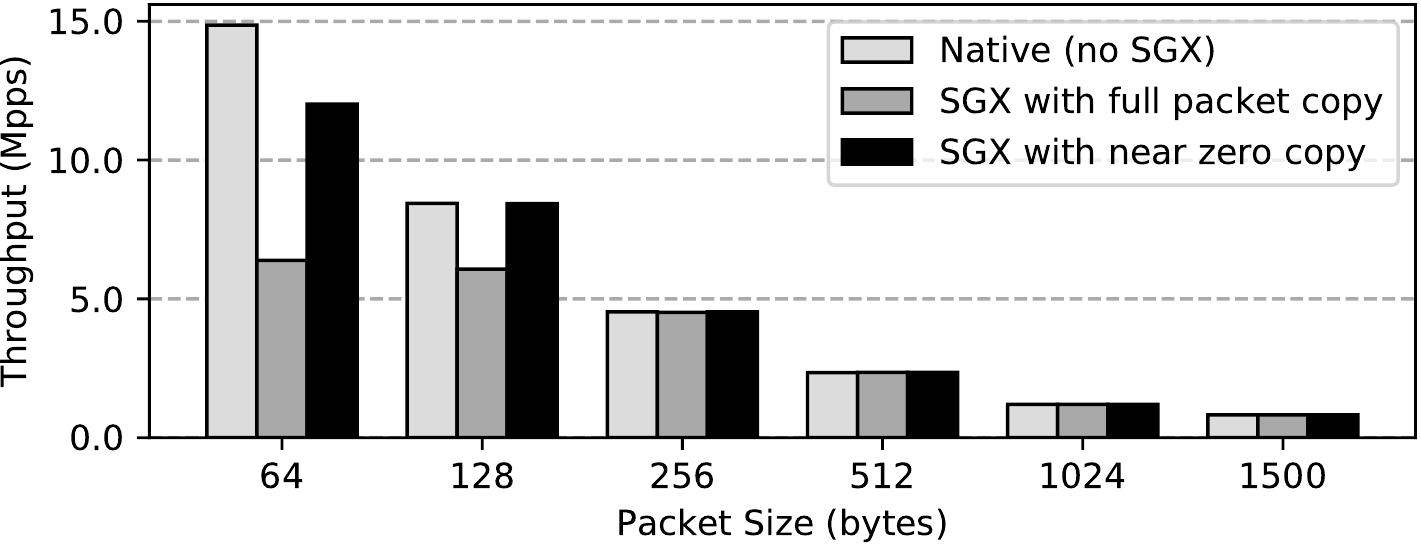}
	\caption{Throughput performance in packet-per-second for varying packet sizes and 3,000 rules with three implementation versions: (1) Native (no SGX), (2) SGX with full packet copy, and (3) SGX with near zero copy.}
	\vspace{-5pt}
	\label{fig:throughput-pps}
\end{figure}

Figure~\ref{fig:throughput-pps} shows the throughput evaluation in packet per second metric. 
Taking a closer look at the SGX with full packet copy implementation, we notice that the maximum packet processing rate is capped at roughly 6 Mpps, which suggests the inherent capacity limit of the full packet-copy operations.
Unlike the full packet copy version, the near zero-copy version shows no such throughput cap in terms of packet per second. 

\section{Connection-Preserving Filtering Performance of \name}
\label{sec:connection-performance}

As we discussed in Appendix~\ref{sec:filter-design-points}, the two mechanisms for connection preservation (i.e., hash-based filtering, exact-match rule based filtering) have different advantages and disadvantages.
Thus, we present a {\em hybrid} design for practical operations.
For any new flow that does not match any existing exact-match filter rules, the filter allows/drops based on the hash digest of the 5-tuple of the packets and queues this 5-tuple.
At every filter rule update (e.g., every 5 seconds)\footnote{The rule update period can be synchronized with that of the rule re-configuration for scalable, multiple enclave operations; see Section~\ref{sec:optimization-multiple-enclaves}.}, all the newly received flows since the last update are converted into exact-match rules and inserted to the lookup table. 
This hybrid design amortizes the cost of lookup table update by batch processing multiple newly observed flows at every update period.
Also, it limits the per-packet latency increase due to hash operations since newly observed flows should be the minority in general. 
Indeed, our experiments on the performance overhead of the use of hash-based filtering  with various packet sizes show no performance degradation, except with small packet size.

\begin{figure}[t!]
	\centering
	\includegraphics[width=0.4\textwidth]{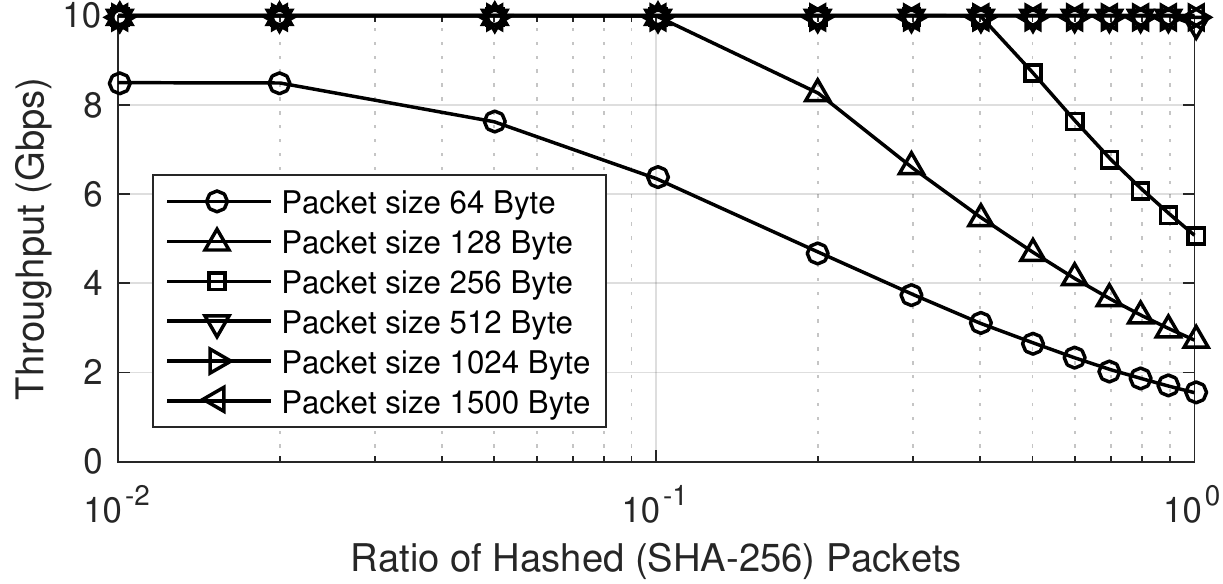}
	\caption{Throughput performance of the maximum 10 Gbps \name filter when a varying fraction of packets are hashed (i.e., SHA-256 is calculated for the 5-tuple bits)}
	\label{fig:hash-performance}
\end{figure}

Figure~\ref{fig:hash-performance} shows our experiment on the performance overhead of the use of hash-based filtering for varying fraction of incoming packets.
Particularly, when the ratio of hashed packets is low (e.g., $< 10$ \%), we observe no performance degradation in all packet sizes, except the smallest size (i.e., 64 Byte) where up to 25\% throughput degradation is measured. 
We argue that this performance degradation is easily acceptable because in general, the fraction of newly observed flows within a short period (e.g., 5 seconds) would be small. 
Moreover, the 64-Byte performance degradation in Figure~\ref{fig:hash-performance} must be the lower-bound result since it assumes that all the packets are 64 Byte short packets. 

\begin{table}[t]
	\centering
	\footnotesize
	\caption{Overhead of filter rule batch insertion to a multi-bit trie lookup table.}
	\label{tab:batch-insertion}
	\begin{tabular}{l r r r r}
		\toprule
		Number of rules in a batch &  1 & 10 & 100 & 1000  \\
		\midrule
		Insert time (millisecond) & 50  &  52 & 53 & 75  \\
		\bottomrule
	\end{tabular}
	\vspace{-15pt}
\end{table}

Table~\ref{tab:batch-insertion} shows the benchmark on the time taken to insert the batched new exact-match rules to a multi-bit tri-based lookup table. 
Our test shows that the batch insertion of filter rules is quite efficient and incurs minimal performance overhead even for large batch size; e.g., only 75 milliseconds compared to the 5-second rule update period. 

\section{Remote Attestation Performance}
\label{app:remote-attestation-performance}
\codename performs remote attestation for each new enclave that the \name IXP
launches on its infrastructure. Since \codename is expected to operate under
DDoS attacks, we want to ensure that the launching of multiple-enclaves
on demand does not become the bottleneck for our deployment model. 
Here, we measure
the total amount of time to complete an end-to-end remote attestation
process for one enclave. In our micro-benchmark for remote attestation for the conservative performance tests, we set up the filter enclave
and the destination on a cloud machine hosted in
South Asia, and the IAS service
hosted in Ashburn, Virginia, United States. For an enclave binary of size $1$
MB, the platform takes $28.8$ milliseconds and the total end-to-end latency of $3.04$ seconds with a standard deviation of $9.2$ milliseconds.

\section{Top Regional IXPs}
\label{sec:top-ixp}

\begin{table*}[ht!]
	\centering
	\footnotesize
	\caption{Top five IXPs in each of the five regions. Numbers inside the parentheses denote the member sizes of the IXPs.}
	\label{tab:top-ixps}
	\begin{tabular}{clllll}
		\toprule
		Rank & Europe &  North America & South America & Asia Pacific & Africa  \\
	    \midrule
		1 & AMS-IX (1660) & Equinix Ashburn (598)  &  IX.br S\~{a}o Paulo (2082) & Equinix Singapore (504) & NAPAfrica Johannesburg (506) \\
		2 & DE-CIX (1494) & Any2 (557)  &  PTT Porto Alegre (258) & Equinix Sydney (393) & NAPAfrica Cape Town (258) \\
		3 & LINX Juniper (755) & SIX (462) &  PTT Rio de Janeiro (246) & Megaport Sydney (383) & JINX (180) \\
		4 & EPIX Katowice (732) & TorIX (426) &  CABASE-BUE (183) & BBIX Tokyo (286) & NAPAfrica Durban (122) \\
		5 & LINX LON1 (697) & Equinix Chicago (384) &  PTT Curitiba (140) & HKIX (281) & IXPN Lagos (69)  \\
		\bottomrule
	\end{tabular}

\end{table*}

We use the IXP peering membership from CAIDA~\cite{caida-ixp} to count the number of AS members of each IXP and summarize the top five IXPs in each of five regions (Europe, North America, South America, Asia Pacific and Africa) in Table~\ref{tab:top-ixps}.

\end{document}